\documentclass[]{article}
\usepackage{authblk}
\usepackage[title]{appendix}
\usepackage{color}
\usepackage[dvipsnames]{xcolor}
\usepackage{amssymb}
\usepackage{amsmath,mathrsfs}
\usepackage{braket}
\usepackage{bm} 

\usepackage[ruled]{algorithm2e}
\usepackage{algorithmic}
\usepackage{setspace,soul,comment}
\usepackage{graphicx}
\usepackage{grffile}
\usepackage{subfigure}
\usepackage{hyperref}

\usepackage{cleveref}
\usepackage[utf8]{inputenc}
\usepackage{geometry}
\newcommand{\dd}{\mathrm{d}}
\usepackage{float}

\title{Inferring epistasis from genomic data with comparable mutation and outcrossing rate}

\author[1,2]{Hong-Li Zeng \thanks{These two authors contributed equally}}
\affil[1]{School of Science, and New Energy Technology Engineering Laboratory of Jiangsu Province, Nanjing University of Posts and Telecommunications, Nanjing 210023, China}
\affil[2]{Nordita, Royal Institute of Technology, and Stockholm University, SE-10691 Stockholm, Sweden}

\author[3]{Eugenio Mauri \thanks{These two authors contributed equally}}
\affil[3]{Laboratory of Physics of the Ecole Normale Sup\'{e}rieure, CNRS UMR 8023 and PSL Research, 24 rue Lhomond, 75231 Paris cedex 05, France}

\author[4,5,6]{Vito Dichio}
\affil[4]{Institut du Cerveau et de la Moelle épinière, ICM, F-75013, Paris, France}
\affil[5]{Inria, Aramis project-team, F-75013, Paris, France}
\affil[6]{Sorbonne Université, F-75013, Paris, France}

\author[3]{Simona Cocco}

\author[3]{R\'emi Monasson}

\author[7]{Erik Aurell}
\affil[7]{KTH -- Royal Institute of Technology, AlbaNova University Center, SE-106 91 Stockholm, Sweden}%

\begin{document}
\maketitle
\begin{abstract}
We consider a population evolving due
to mutation, selection and recombination, where selection includes single-locus terms (additive fitness) and
 two-loci terms (pairwise epistatic fitness).
We further consider the problem of
inferring fitness in the evolutionary dynamics
from one or several snap-shots of the
distribution of genotypes in the population.
In the recent literature this
has been done by applying the
Quasi-Linkage Equilibrium (QLE) regime first obtained by Kimura in the limit of high recombination.
Here we show that the approach also works
in the interesting regime where the effects of mutations are comparable to or larger than recombination.
This leads to a modified main epistatic fitness
inference formula where the rates of mutation
and recombination occur together.
We also derive this formula
using by a previously developed
Gaussian closure that formally
remains valid
when recombination is absent.
The findings are validated through numerical simulations.
\end{abstract}


\textbf{Email}: hlzeng$@$njupt.edu.cn; emauri$@$clipper.ens.psl.eu; vito.dichio$@$etu.sorbonne-universite.fr;\\
\indent ~~~~~~~~~~ simona.cocco$@$phys.ens.fr;  monasson$@$lpt.ens.fr; eaurell$@$kth.se

\section{Introduction}
\label{sec:introduction}
Fitness as understood in this paper
is the propensity of an organism to pass on
its genotype to the next generation,
described by a fitness value of
each genotype. A set of such values
is called a fitness landscape;
evolution is a process whereby nature
tends towards populating the peaks in the landscape~\cite{deVisser2014}.
Motion in fitness landscapes
describes the evolution of a population
of one species
in a roughly
constant environment.
Prime examples of this are
pathogens and parasites colonizing
a host evolving on a much slower time scale.
The most fit pathogen is then one
that is best able to exploit the opportunities
and weaknesses of a typical host to grow,
multiply and eventually spread to other hosts.
Excluded from the concept of fitness as considered
here are aspects of games of competition
and cooperation in evolution~\cite{MaynardSmith82,Chastain2014}.

Sequencing of genomes of human
pathogens today happen on a massive
scale. In an extreme example, samples of
SARS-CoV-2, the etiological agent of
the disease COVID-19,
have by now been sequenced more than
1,200,000 (accessed on 2021-04-23)
times, and is being sequenced many thousands of times daily~\cite{GISAID,Nextstrain,Hadfield2018}.
This virus in the betacoronavirus family
has only been known to science for about 16 months.

It is clear that much information
about the evolutionary process must be contained
in such data.
In particular, if genetic variants in different
positions contribute synergistically to fitness
this should be reflected in the distribution over
genotypes.
The goal of this paper is to address the
basis of such an approach, and to develop tools to
use it better in the future.
In two recent
contributions~\cite{Gao2019,Zeng2020} we have
argued that a natural setting is the
\textit{Quasi-Linkage Equilibrium} (QLE)
phase of
Kimura~\cite{Kimura1965}, surveyed by Kirkpatrick, Johnson and  Barton \cite{Kirkpatrick2002}, and 
more recently studied by Neher and
Shraiman~\cite{NeherShraiman2009,NeherShraiman2011}.
When recombination
(the exchange of genomic material between individuals, or sex)
is a much faster process than mutations or selection due to fitness
the stationary distribution over genotypes
is the Gibbs-Boltzmann distribution of
an Ising or Potts model.
The inverse Ising/Potts~\cite{Roudi-2009b,Nguyen-2017a} or Direct Coupling Analysis (DCA)~\cite{Morcos-2011a,Stein-2015a,Cocco-2018a}
methods have been invented to infer the parameters
of such distributions from samples.
Quantitative properties of QLE allow to go one step
further, and relate those effective couplings
to the parameters of the evolutionary dynamics,
which we will call the Kimura-Neher-Shraiman (KNS) theory.
In~\cite{Zeng2020} we showed that it is indeed possible
to retrieve synergistic contributions to fitness
from simulated population data by KNS theory.

In the following we will present
an extension where we relax the
requirement that recombination has to
be the fastest process in the problem.
Instead we allow for either recombination
or mutation being the fastest process
and derive a new modified epistatic
inference formula.
We do this both by adapting the
argument from QLE~\cite{NeherShraiman2011}
and by
a Gaussian closure recently
developed by three of us~\cite{Mauri-2019,Mauri-2020}.
We will show that this new theory
allows for retrieving synergistic contributions to fitness
in much wider parameter ranges.
Recombination (sex) is hence no longer required to be a much stronger process than mutations, but could
in the Gaussian closure
actually be set to zero.
The conditions
on recombination
compared to variations in synergistic contributions to fitness
are also much less strict in the new theory.

The paper is organized as follows.
In Section~\ref{sec:evolutionary-dynamics}
we summarize evolution driven by
selection, recombination and mutations,
and contrast the different
epistasis inference formulae.
In Section~\ref{sec:QLE-low-recombination}
we derive the formula at high mutation
but not necessarily high recombination
within QLE,
while in Section~\ref{sec:Gaussian-closure}
we do it from the Gaussian closure ansatz.
In Section~\ref{sec:simulation}
we summarize our model and simulation strategies,
and in Sections~\ref{sec:mutation-vs-recombination}
and~\ref{sec:fitness-vs-recombination}
we compare how well we are able to
infer fitness when varying mutation rate,
the strength of fitness variations, and the rate
of recombination.
In Section~\ref{sec:discussion}
we summarize and discuss our results.
Appendices contain additional material.
Appendix \ref{a:Higher-order-GC} computes higher order terms for the inference formula in the Gaussian closure scheme.
Appendix~\ref{a:FFPopSim} contains parameter settings for
simulations of an evolving population
using the FFPopSim software~\cite{FFPopSim},
and in Appendix~\ref{a:nMF}
we give details on the DCA method we
have used in this work. Appendix \ref{a:fitness-dca-chi} presents the comparisons of equations obtained from QLE and Gaussian closure. 
Appendix \ref{a:Genetic Drift} shows the effects of genetic drift on the epistasis inference. Appendix \ref{a:Gaussian_additive_effects} provides the epistasis inference with Gaussian distributed additive effects. 

\section{Evolutionary dynamics and epistasis inference}
\label{sec:evolutionary-dynamics}
The forces of evolution in classical population
genetics are selection, mutations and genetic drift
\cite{Fisher-book,Blythe2007}.
Selection confers an advantage
on individuals with certain characteristics,
so that they tend to have more descendants.
Mutations are random changes of the genomes.
Genetic drift is the element of chance
as to which individual survives, and which does not.
Common to these three forces is that they all act
on the single genotype level: an organism survives
to the next generation, or it does not.
If it does
it will have a number of descendants ``children'', ``grand-children'' etc.
The distribution of individuals over genotypes
can then formally be written as a gain-loss process
\begin{equation}
\label{eq:gain-loss}
\partial_t P(\mathbf{g},t) = \sum_{\mathbf{g}'}\left(k_{\mathbf{g}',\mathbf{g}} P(\mathbf{g}',t)-k_{\mathbf{g},\mathbf{g}'} P(\mathbf{g},t)\right),
\end{equation}
where the rates $k_{\mathbf{g}',\mathbf{g}}$ encode selection and mutation. Genetic drift can not be described by eq.(\ref{eq:gain-loss}) directly, which is valid in the infinite population size limit, but appears in Monte Carlo simulation naturally through finite $N$ effects.
The details of relevant equations are discussed in great detail
in~\cite{NeherShraiman2011} as well as more
recently in~\cite{Gao2019,Zeng2020}.

Recombination (or sex) is the process by which two genotypes
combine to give a third one in the next generation. It cannot be
expressed in the form of eq. \eqref{eq:gain-loss}.
Instead, in general terms it looks as
\begin{equation}
\partial_t P(\mathbf{g},t) = \cdots +
\sum_{\mathbf{g}',\mathbf{g}''} C_{\mathbf{g},\mathbf{g}',\mathbf{g}''} P_2(\mathbf{g}',\mathbf{g}'',t),
\label{eq:collision}
\end{equation}
where $P_2$ stands for the joint probability of two genotypes
$\mathbf{g}'$ and $\mathbf{g}''$,
and $C_{\mathbf{g},\mathbf{g}',\mathbf{g}''}$ is the rate at which these two produce an offspring
$\mathbf{g}$.
Equation~\eqref{eq:collision} is not closed; there would be
an equation for $\partial_t P_2$ which would depend on the
three-genotype distribution $P_3$, and so on.
A standard way to close such a BBKGY-like hierarchy is to
assume random mating (random collisions), \textit{i.e.},
$P_2(\mathbf{g}',\mathbf{g}'',t)=P(\mathbf{g}',t)P(\mathbf{g}'',t)$.
Combining \eqref{eq:gain-loss} and \eqref{eq:collision}
we hence get the evolution of a population as a non-linear differential
equation analogous to a Boltzmann equation.

In \eqref{eq:gain-loss} and \eqref{eq:collision}
each genotype $\textbf{g}$ is seen as a sequence of positions (or loci) of length $L$, $\textbf{g}\equiv \{s_0,s_1,\dots, s_{L-1}\}$. The variable at each position (the allele) $s_i$ can be in one out of $n_i$ states. In the following discussion, we simplify by taking $n_i = 2$ such that $s_i$ is a binary variable. Following the conventions in the physical literature, and in particular~\cite{NeherShraiman2011}, we set $s_i = \pm 1$.

We will from now on limit ourselves to fitness landscapes
that contain linear and quadratic terms in the allele
variables. This means that the fitness
of a genotype is given by a function
\begin{equation}
  F(\mathbf{g}) = \sum_i f_i s_i + \sum_{ij} f_{ij} s_i s_j
\label{eq:fitnessapprox}
\end{equation}
The linear term $f_i$ is called an
\textit{additive contribution to fitness}
while the quadratic $f_{ij}$ is an
\textit{epistatic contribution to fitness}.
The goal of the line of research
pursued in this paper is to find ways to
retrieve the $f_{ij}$ from the distribution
of genotypes in a population.

The Quasi-Linkage Equilibrium (QLE) theory is
based on approximating the genome distribution $P$ as
a Gibbs-Boltzmann distribution of the Ising/Potts type:
\begin{equation}
    \log P(\textbf{g},t) = \Phi(t) + \sum_i\phi_i(t) s_i + \sum_{i<j}J_{ij}(t) s_is_j ,
    \label{eq:Boltz_ditr}
\end{equation}
In above $\Phi(t)$ is a normalization factor
playing the same role as
$-\beta F(\beta)$ in statistical mechanics.
By expressing the evolution equations
for $P(\textbf{g},t)$ in terms of the effective
parameters $\Phi(t)$, $\phi_i(t)$
and $J_{ij}(t)$,
it was shown
in~\cite{Kimura1965,NeherShraiman2009,NeherShraiman2011},
that the distribution \eqref{eq:Boltz_ditr}
is stable at high rate of recombination.
The values of the parameters
 $\Phi$, $\phi_i$
and $J_{ij}$ in stationary state are
then related to the model parameters
as discussed in detail in~\cite{NeherShraiman2011,Gao2019}.
In particular, $J_{ij}$ is simply proportional
to $f_{ij}$
which can be turned around
to the \textit{KNS fitness inference formula}
\begin{equation}
  f^*_{ij} = J^*_{ij} \cdot rc_{ij}.
  \label{eq:original_KNS_equation}
\end{equation}
The stars on both sides indicate that these
are inferred quantities,
the proportionality
parameters $r$ and $c_{ij}$ are discussed below.



In this work we will extend the above analysis
to the regime where recombination is not
necessarily high, but mutation remains
a faster process than selection.
In Section~\ref{sec:QLE-low-recombination}
we derive this within QLE,
and in Section~\ref{sec:Gaussian-closure}
we do it by Gaussian closure.
Here we discuss and contrast these
different (though related) formulae.

In QLE with mutation comparable to or larger than
recombination the relevant inference formula changes to
\begin{equation}
  f^*_{ij} = J^*_{ij} \cdot \left(4\mu + rc_{ij}\right),
  \label{eq:low-recomb-_KNS_equation}
\end{equation}
where $\mu$ is the rate of mutations assumed to be the
same at all loci and in both directions.
In both \eqref{eq:original_KNS_equation}
and  \eqref{eq:low-recomb-_KNS_equation}
the Gibbs-Boltzmann parameter  $J^*_{ij}$
is not directly observed, but has to be inferred
from the data.
All such procedures, collectively known either
as inverse Ising/Potts or as Direct Coupling Analysis,
have to make a trade-off between accuracy and computability.
Let us here mention the benchmark statistical
method of \textit{maximum likelihood}
which is accurate but not efficiently computable
in large systems, and \textit{naive mean-field inference}
(described in Appendix~\ref{a:nMF}) which
amounts to matrix inversion of the empirical
correlation matrix.
Other procedures were reviewed in~\cite{Roudi-2009b,Nguyen-2017a},
see~\cite{Morcos-2011a,Stein-2015a,Cocco-2018a}.
A particular DCA procedure introduced in~\cite{NeherShraiman2011}
is \textit{small interaction expansion} (SIE)
\begin{equation}
\label{eq:SIA}
J^{*,SIE}_{ij} = \chi_{ij} /\left(\left(1-\chi_i^2\right)\left(1-\chi_j^2\right)\right)
\end{equation}
where * stands for the type of inference used,
and $\chi_i \equiv \braket{s_i}$ and
$\chi_{ij} \equiv \braket{s_is_j} - \chi_i\chi_j$
are the (connected) first and second order correlation functions.
Inference formula \eqref{eq:SIA} is not very accurate as a general DCA
method~\cite{NeherShraiman2011,Gao2019},
but has the advantage of being eminently computable.
Substituting \eqref{eq:SIA} in \eqref{eq:low-recomb-_KNS_equation}
one gets
\begin{equation}
  f^{*,SIE}_{ij} = \frac{\chi_{ij}}{\left(\left(1-\chi_i^2\right)\left(1-\chi_j^2\right)\right)} \cdot \left(4\mu + rc_{ij}\right)
  \label{eq:fitness-inference-formula}
\end{equation}
As it will turn out, \eqref{eq:fitness-inference-formula}
is also the formula which appears directly in Gaussian
closure. We can hence derive \eqref{eq:fitness-inference-formula}
in two different ways.

In all above, the parameters $\mu$, $r$ and $c_{ij}$
have the same meaning as in~\cite{NeherShraiman2011}
and stand for mutation rate (assumed uniform),
recombination rate (assumed uniform)
and the probability of off-springs inheriting the genetic information from different parents.  For high-recombination organisms, $c_{ij}$ depends on the cross-over rate $\rho$ and the
genomic distance between loci $i$ and $j$ \cite{Zeng2020}, except when loci $i$ and $j$ are very closely spaced on the genome.
\begin{equation}
c_{ij}\approx \frac{1}{2}\left(1-e^{-2\rho|i-j|}\right)
\label{eq:cij}
\end{equation}

When comparing
  \eqref{eq:original_KNS_equation} and
  \eqref{eq:fitness-inference-formula}
in numerical testing we simulate an evolving population
at the same parameter values, and then either use
the genotype information to compute
empirical correlations, or to infer
Ising/Potts parameters by DCA.
For simplicity we will in the following
only present results obtained by DCA naive mean-field (nMF)
inference; results are very similar
for other common variants of DCA.

\section{Quasi-linkage equilibrium outside high recombination}
\label{sec:QLE-low-recombination}

In this Section we introduce the model defined in~\cite{NeherShraiman2011} for the evolution of the distribution of genomes $P(\textbf{g},t)$ and discuss the high mutation regime in order to recover the inference formula for epistatic interactions \eqref{eq:low-recomb-_KNS_equation}. Throughout we
assume an infinite population; genetic drift is therefore not considered.

Selection is the first fundamental ingredient and works as follows: each possible sequence $\textbf{g}$ grows inside the population with a certain growth-rate $F(\textbf{g})$, called \textit{fitness}, which can be described as a function of the specific sequence $\textbf{g}$. As stated above in eq. \eqref{eq:fitnessapprox}, we will approximate any fitness function $F$ as the sum of linear terms $f_i$, called \textit{additive fitness}, and pairwise iteractions $f_{ij}$, called \textit{epistatic fitness}. Note that in general one can also include higher order terms of the form $f_{i_1 ... i_n}s_{i_1}...s_{i_n}$.

	
The second ingredient for the population evolution is mutations. We assume that in each small time interval $\Delta t \ll 1$ a fraction $\mu\Delta t$ of all the alleles inside the population ($L$ for each individual) mutate by a single spin-flip; $\mu$ is therefore named mutation rate. We describe the process of a spin flip by introducing an operator $M_i$ acting on a sequence by changing the sign of the $i$-th spin. To understand how the frequency of a certain sequence $\textbf{g}$ changes in the interval $\Delta t$, we should count how many individuals have mutated into the sequence $\textbf{g}$ and how many sequences have instead mutated from away this state.
	
The last element to consider is recombination between different sequences. At each small time interval $\Delta t$ a fraction  $r\Delta t $ of the individuals (where $r$ is the recombination rate) encounters random pairing and crossing-over, giving rise to new genomes. The evolution of the distribution $P$ in the interval $\Delta t$ due to recombination is given by
\begin{equation}
    P(\textbf{g}, t+\Delta t)= (1-r \Delta t)P(\textbf{g},t)+ r\Delta t\!\!\! \sum_{ \{s'_i\} \{\xi_i\}}\!\!\!C(\{\xi\}) P(\textbf{g}^{(m)},t)P(\textbf{g}^{(f)},t)\,.
	\label{eq:recombination}
\end{equation}
The first term counts for those individuals that did not recombine during the time interval $\Delta t$. When two individuals recombine, a new genotype is formed by inheriting some loci from the the mother with genotype $\textbf{g}^{(m)}$ and the complement from the father with genotype $\textbf{g}^{(f)}$. The parts of the genomes of the mother and the father not inherited by the child (and hence discarded) is denoted $\textbf{g}^{'}$. 
The cross-over can be described by a vector $\{\xi_i\}$, with $\xi_i \in \{0,1\}$. If $\xi_i = 1$, the $i$-th locus is inherited from the mother, otherwise from the father.
Turning around the relation we have $s_i^{(m)} = s_i\xi_i + s_i'(1-\xi_i)$ and  $s_i^{(f)} = s'_i\xi_i + s_i(1-\xi_i)$ where $s_i$ is the allele of the child at locus $i$, and $s_i'$ is the discarded allele.
The probability of each realization of $\{\xi_i\}$ is given by $C(\{\xi\})$. Subsequently we need to sum over all the possible genomes which are not passed on the offspring
($\textbf{g}^{'}$) as well as all the possible crossover patterns $\{\xi\}$~\cite{NeherShraiman2011,Gao2019}.

Merging together all the ingredients, we obtain the following non-linear differential equation for the time derivative of the genotype distribution $P$:
\begin{equation}
    \begin{split}
        \dot{P}(\textbf{g},t)  &=\frac{\dd }{\dd t}|_{\text{fitness}}\, P(\textbf{g},t) + \frac{\dd }{\dd t}|_{\text{mut}}\, P(\textbf{g},t) +\frac{\dd }{\dd t}|_{\text{rec}}\, P(\textbf{g},t)  \\
        & =  \left[F(s)-\braket{F}\right] P(\textbf{g},t) +  \mu \sum_{i=0}^{L-1}\left[ P(M_i\textbf{g},t)- P(\textbf{g},t) \right] +\\
        &+ r\!\!\!\!\sum_{ \{s'_i\} \{\xi_i\}}\!\!\!\!C(\{\xi\})\left[  P(\textbf{g}^{(m)},t)P(\textbf{g}^{(f)},t) - P(\textbf{g},t)P(\textbf{g}',t) \right]\,.
    \end{split}{}
    \label{eq:master_eq}
\end{equation}{}

Now we want to study the stationary solutions of this master equation. In particular, we seek to extend Neher and Shraiman's argument \cite{NeherShraiman2011} in the {high mutation} limit, recovering the inference formula \eqref{eq:low-recomb-_KNS_equation} of the epistatic interactions introduced above. We start by assuming the distribution $P$ to be of the same form as in eq.~\eqref{eq:Boltz_ditr}:
	\begin{equation}
		P(g,t) = \frac{1}{Z(t)}\exp\left[\sum_{i}\phi_i(t)s_i + \sum_{i<j}J_{ij}(t)s_is_j \right]\,,
	\end{equation}
	where $Z(t)$ is the normalization factor. Following Neher and Shraimann in \cite{NeherShraiman2011}, we now inject this ansatz in the master equation for the evolution of $\log P(g,t)$ in presence of mutations and recombination and obtain
	\begin{align}
	     \frac{\dd \log P(g,t)}{\dd t}&=-\frac{\dd}{\dd t}\log Z(t) + \sum_i\dot\phi_i(t)s_i +\sum_{i<j}\dot J_{ij}(t)s_is_j \nonumber\\ 
	     &= F(g) - \langle F \rangle + \mu\underbrace{\sum_i \left[ \frac{P(M_i g,t)}{P(g,t)} - 1\right]}_{M(g,t)} + r\underbrace{\sum_{ \{\xi_i\}\{s^{'}_i\}}C(\{\xi\})P(g',t)\left[\frac{P(g^{(m)},t)P(g^{(f)},t)}{P(g,t)P(g^{'},t)}-1\right]}_{R(g,t)}\,.
	     \label{eq:logPevo}
	\end{align}
	Now we separate the mutation and recombination term ($M(g,t)$ and $R(g,t)$, respectively) from the RHS of the last equation and compute them separately. Starting from the recombination term, we may rewrite it as
	\begin{equation}
	    R(g,t) = \sum_{ \{\xi_i\}\{s^{'}_i\}}C(\{\xi\})P(g',t)\left[e^{\sum_{i<j}J_{ij}\left[(\xi_i\xi_j+\overline{\xi}_i\overline{\xi}_j-1)(s_is_j+s'_is'_j)+(\xi_i\overline{\xi}_j+\overline{\xi}_i\xi_j)(s_is'_j+s_is'_j)\right]}-1\right]\,,
	\end{equation}
	where $\overline{\xi}_i=(1-\xi_i)$. Now in the high recombination limit considered in \cite{NeherShraiman2011} the authors suppose that the interactions $J_{ij}$ are small and can be expanded from the exponential. We note that the same argument should hold also when mutations are dominant in the evolution. Hence, we write
	\begin{align}
	    R(g,t) &\sim \sum_{ \{\xi_i\}\{s^{'}_i\}}C(\{\xi\})P(g',t)\left[{\sum_{i<j}J_{ij}\left[(\xi_i\xi_j+\overline{\xi}_i\overline{\xi}_j-1)(s_is_j+s'_is'_j)+(\xi_i\overline{\xi}_j+\overline{\xi}_i\xi_j)(s_is'_j+s_is'_j)\right]}\right] \nonumber \\
	    &= \sum_{i<j}c_{ij}J_{ij}\left[(s_i\langle s_j \rangle +s_j\langle s_i \rangle ) - (s_is_j + \langle s_is_j \rangle)\right]\,,
	\end{align}
	where $c_{ij} \equiv \sum_{{\xi}}C(\{ \xi \})\left[\xi_i\overline{\xi}_j + \overline{\xi}_i\xi_j\right]$ represents the probability that loci $i$ and $j$ arrive from different parents. 
	
	Now we turn to the mutation term $M(g,t)$ that can be written as follows:
	\begin{equation}
	    M(g,t) = \sum_i \left[ e^{-2\phi_is_i -2\sum_j J_{ij}s_is_j} - 1\right]\,.
	\end{equation}
	In the high mutation regime we suppose that both the interactions and the fields are small and can be expanded from the exponential:
	\begin{equation}
	    M(g,t) \sim -2\sum_i \phi_is_i  -4\sum_{i<j} J_{ij}s_is_j \,.
	\end{equation}
	Injecting these results for $M(g,t)$ and $R(g,t)$ into eq.~\eqref{eq:logPevo} and separating the dependencies on $s_i$ and $s_is_j$, we can obtain equations for $\dot{\phi}_i$ and $\dot{J}_{ij}$ similarly to what has been done by Neher and Shraimann in \cite{NeherShraiman2011}. In particular, we find:
	\begin{align}
		\dot{\phi}_i &= f_i + r\sum_{j\neq i}c_{ij}J_{ij}\langle s_j \rangle-2\mu\phi_i\\
		\dot{J}_{ij} &= f_{ij} - (4\mu + r c_{ij})J_{ij}\,.
	\end{align}
	Hence, the interactions $J_{ij}$ will quickly evolve through the stationary solution $J_{ij}^{st.}=f_{ij}/(4\mu+rc_{ij})$. Inverting the latter equation we recover the inference formula \eqref{eq:low-recomb-_KNS_equation}.

\section{The argument by Gaussian closure}
\label{sec:Gaussian-closure}

Going forward, we want to parameterize the distribution $P(\textbf{g},t)$ by its cumulants. In particular, we define the cumulants of first and second order as $\chi_i\equiv\braket{s_i}$ and $\chi_{ij}\equiv\braket{s_is_j}-\chi_i\chi_j$. Note that in this way $\chi_{ii}=1-\chi_i^2$. Using eq. \eqref{eq:master_eq}, we can write the time evolution for these cumulants as follows:
    \begin{align}
    \dot{\chi}_i &= \braket{s_i[F(\textbf{g})-\braket{F}]} - 2 \mu \chi_i \label{eq:allele_freq}\\
	\dot{\chi}_{ij} &=\braket{(s_i-\chi_i)(s_j-\chi_j)[F(\textbf{g})-\braket{F}]}-(4\mu + rc_{ij})\chi_{ij} \label{eq:allele_corr}\, ,
    \end{align}
    with $i\neq j$ in the second line and $c_{ij}$ defined as eq.~\eqref{eq:cij}.

In general, eq. \eqref{eq:allele_freq}-\eqref{eq:allele_corr} are not a closed set of equations since they would also depend on higher order cumulants $\chi_{ijk}$, $\chi_{ijkl}$, etc.
The Gaussian closure which we introduced recently~\cite{Mauri-2020} aims to overcome this problem by neglecting those higher order cumulants (connected correlation functions) under the assumption that at high recombination and/or mutations rate their influence on the global dynamics is weak.
For a Gaussian distribution, all cumulants of order higher than two vanish.
With this approximation, \eqref{eq:allele_freq} and \eqref{eq:allele_corr} define a closed set of $L(L+1)/2$ dynamical equations only depending on $\chi_i$ and $\chi_{ij}$.
\begin{alignat}{2}
    \dot\chi_i =& \sum_j\chi_{ij}\Big(f_j+\sum_kf_{jk}\chi_k-2f_{ij}\chi_i\Big) -2\mu\chi_i\label{eq:DFOC}\\
    \dot\chi_{ij}
    =& -2\chi_{ij}\sum_k\Big[f_{ik}(\chi_{ik}+\chi_i\chi_k) +f_{jk}(\chi_{jk}+\chi_j\chi_k)\Big]\ + 2f_{ij}\chi_{ij}(\chi_{ij}+2\chi_i\chi_j)\ + \sum_{k,l}f_{kl}\chi_{ik}\chi_{jl}\ +\notag \\
    & -(4\mu + rc_{ij})\chi_{ij} - 2\chi_{ij}(f_i\chi_i + f_j\chi_j)
 \label{eq:DSOC}
\end{alignat}
In principle, eq. \eqref{eq:DFOC}-\eqref{eq:DSOC} could be simultaneously solved in order to determine the stationary state, which is of our interest, and this in turn would allow to determine the $L(L+1)/2$ quantities $\{f_i\}, \{f_{ij}\}$ as a function of the $\{\chi_i\}, \{\chi_{ij}\}$. Unfortunately, considering the size of the system, this is analytically not feasible.\\

Nevertheless, eq. \eqref{eq:DSOC} suggests another route to infer $f_{ij}$ according to the following argument: when studying the stationary state, we can assume self-consistently that all the off-diagonal $\chi_{ij}$ are small, so that we can expand $\chi_{ij}$, with $i\neq j$, as a power series of $1/(4\mu+rc_{ij})$:
	\begin{equation}
	    \chi_{ij} = \frac{\chi^{(1)}_{ij}}{4\mu+rc_{ij}} + \mathcal{O}((4\mu+rc_{ij})^{-2})\ .
	\end{equation}
	Inserting this in eq.\eqref{eq:DSOC}, we obtain
	\begin{equation}
	        \chi^{(1)}_{ij} = f_{ij} (1-\chi_i^2)(1-\chi_j^2)\ .
	\end{equation}
We therefore conclude that, to the first order,
	\begin{equation}
	    \chi_{ij} = \frac{f_{ij}}{4\mu+rc_{ij}}(1-\chi_i^2)(1-\chi_j^2)\ .
	\end{equation}
Turning around this into an inference formula for fitness we arrive at
\eqref{eq:fitness-inference-formula}.\\


\section{Simulation strategies and results}
\label{sec:simulation}
The basic idea is to simulate the states of a population
with $N$ individuals (genome sequences) evolving under mutation,
selection and recombination and genetic drift.
As in previous work we have used the
FFPopSim package developed by
Zanini and Neher for this purpose~\cite{FFPopSim}.
Simulation and parameter settings
are given in Appendix~\ref{a:FFPopSim}.

In a QLE phase the outcomes of such simulations are trajectories of means $\chi_i(t)$ and correlations $\chi_{ij}(t)$ which in principle can be computed from the configuration of the population $\mathbf{g}^{(\boldsymbol{s})}(t)$ at generation $t$.
After a suitable relaxation period
we take the set $\mathbf{g}^{(\boldsymbol{s})}(t)$
to be independent samples
from a distribution \eqref{eq:Boltz_ditr}
with unknown direct couplings $J_{ij}$.
We will throughout use the DCA algorithm
naive mean-field (nMF) \cite{Kappen-1998a}
to infer parameters $J_{ij}$ from data for original KNS,
for descriptions, see Appendices~\ref{a:nMF}.


The principle of the numerical testing is to
infer epistatic fitness parameters from the
data by \eqref{eq:original_KNS_equation} and
\eqref{eq:fitness-inference-formula},
and then compare to the underlying parameters $f_{ij}$ used to generate the data.
Here, the testing epistatic fitness is Sherrington-Kirkpatrick model \cite{Sherrington-1975} with different variations. The additive fitness $f_i$ follows Gaussian distribution with zero means and the standard deviation $\sigma(\{f_i\})=0.05$ in our simulations.
We note that
\eqref{eq:original_KNS_equation} is proposed to be hold for weak selection and high recombination, and has already been tested in~\cite{Zeng2020}.
Data availability is an issue.
As in~\cite{Zeng2020} we
have used \textit{all-time} versions
of the algorithms, where samples
$\textbf{g}^{(s)}(t)$ at different $t$
are pooled. This is primarily to mitigate the
effect that in a real-world population
the number of individuals $N$ is very large,
but in the simulations it is only moderately
large. All DCA methods as well as empirical
correlations can be more accurately estimated with more samples.

\subsection{Mutation vs recombination rate}
\label{sec:mutation-vs-recombination}
\begin{figure*}[!ht]
\centering
\subfigure{
\begin{minipage}[t]{0.30\linewidth}
\centering
\includegraphics[width=\textwidth]{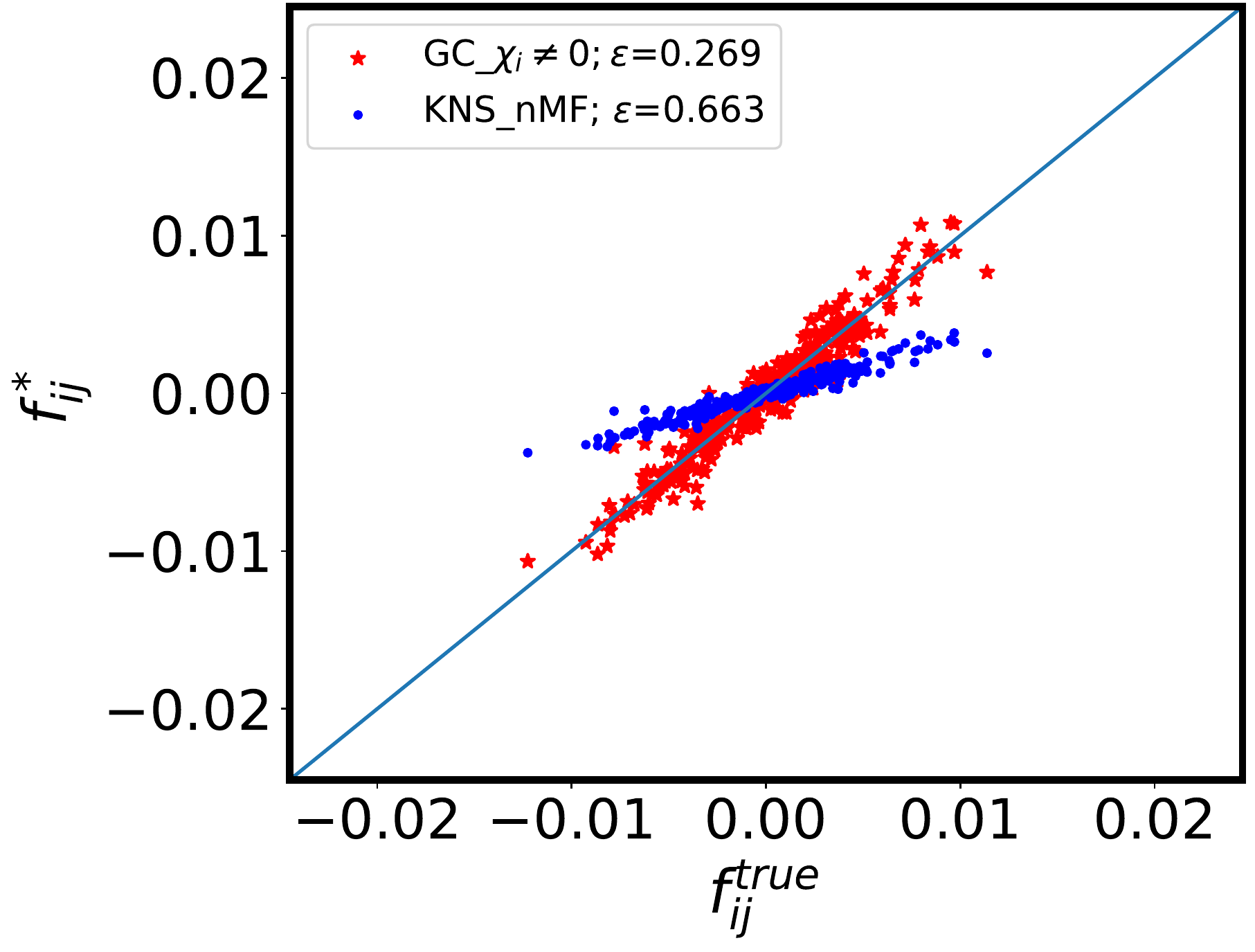}
\put(-101,-11){\small (a)~$\mu=0.05, r=0.2$}
\end{minipage}%
}
\subfigure{
\begin{minipage}[t]{0.3\linewidth}
\centering
\includegraphics[width=\textwidth]{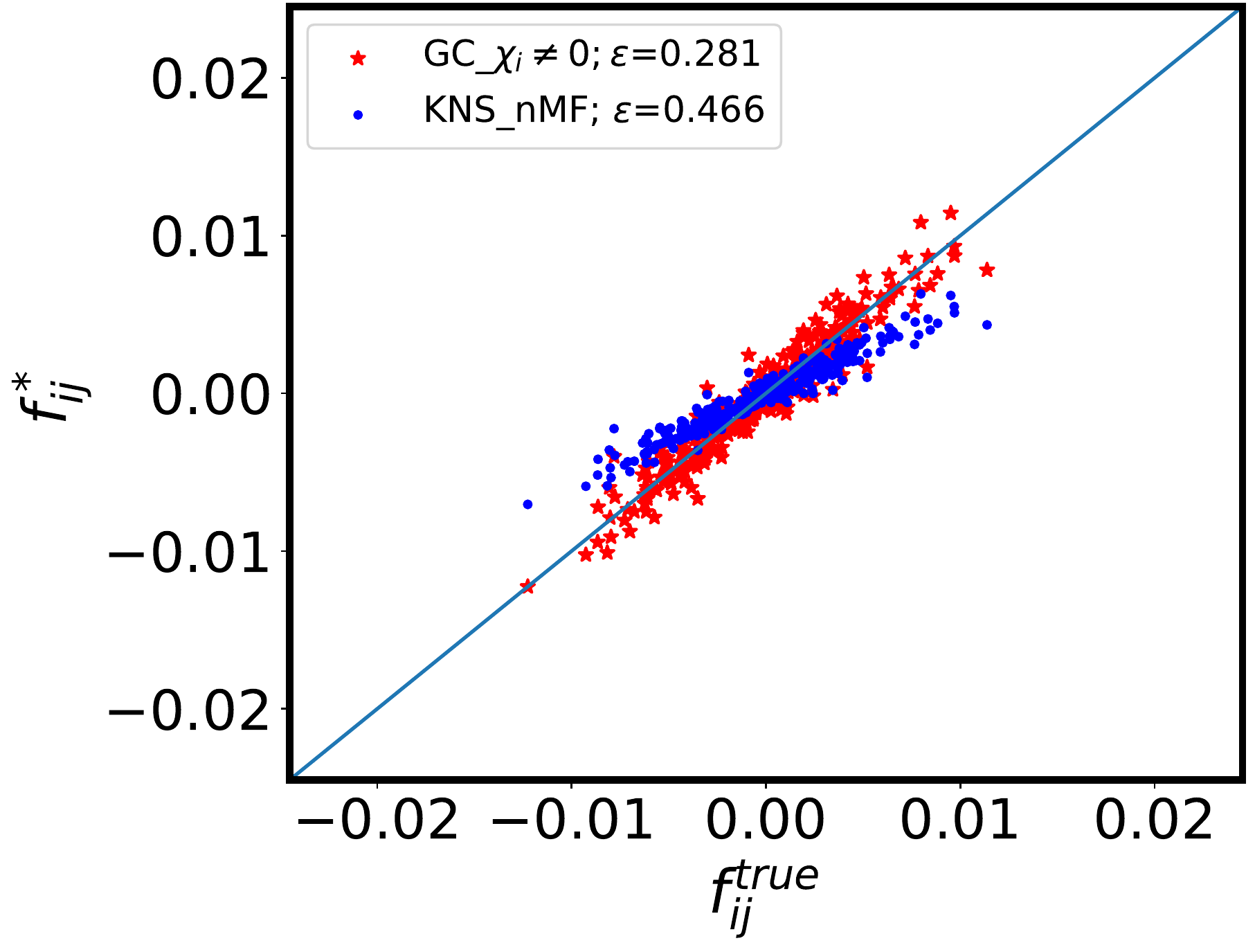}
\put(-101,-11){\small (b)~$\mu=0.05, r=0.5$}
\end{minipage}%
}
\subfigure{
\begin{minipage}[t]{0.3\linewidth}
\centering
\includegraphics[width=\textwidth]{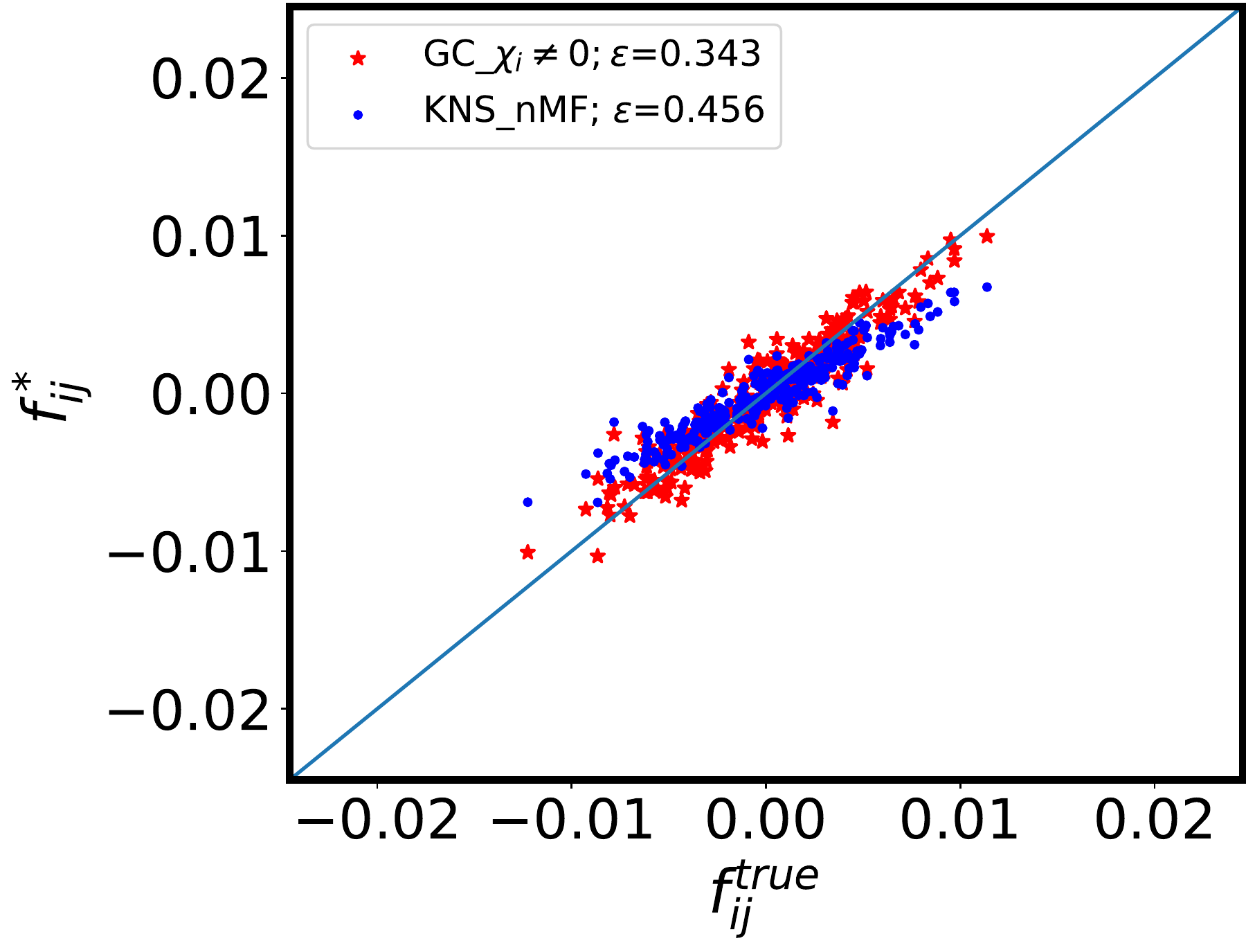}
\put(-101,-11){\small (c)~$\mu=0.05, r=0.8$}
\end{minipage}%
}\\
\subfigure{
\begin{minipage}[t]{0.3\linewidth}
\centering
\includegraphics[width=\textwidth]{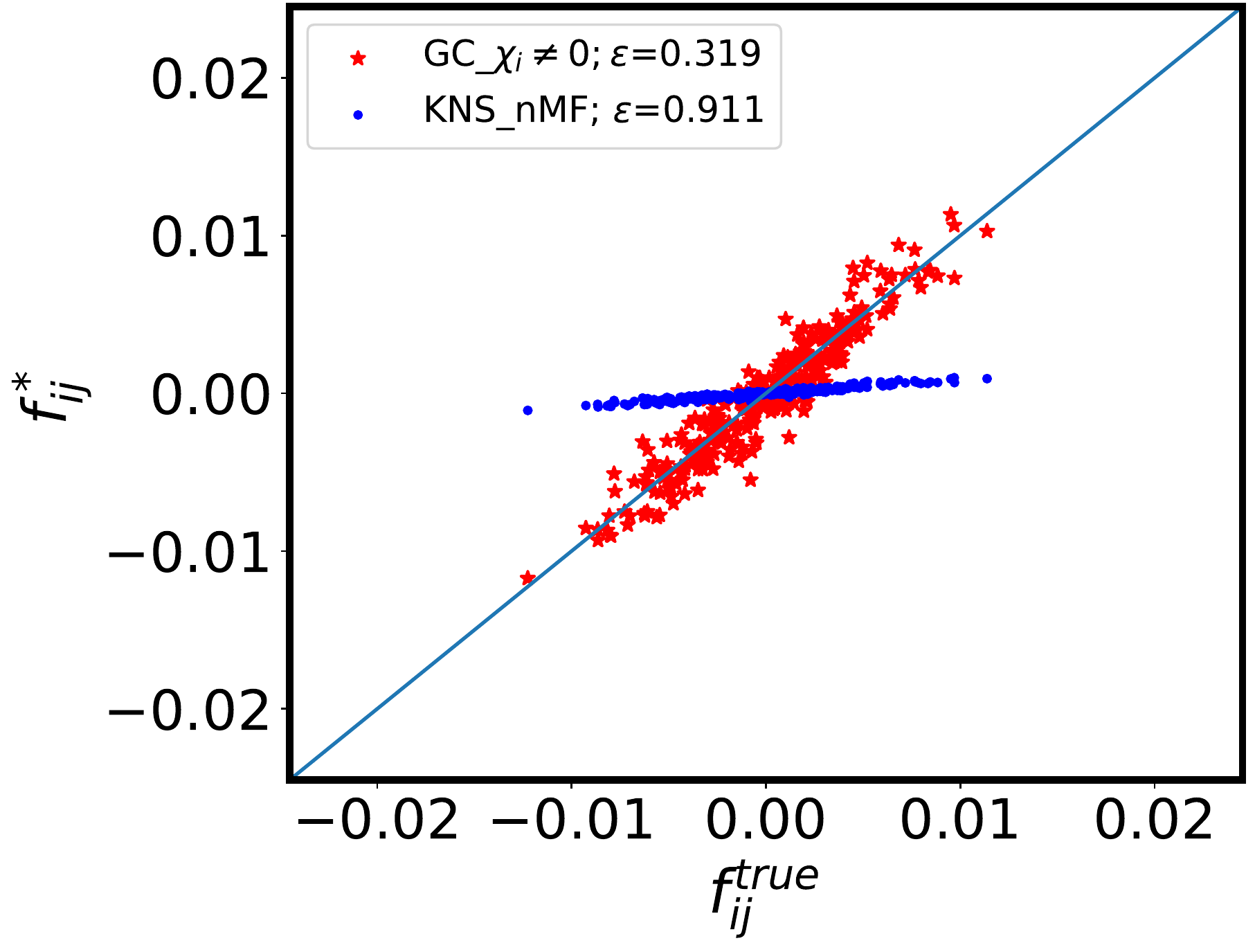}
\put(-101,-11){\small (d)~$\mu=0.25, r=0.2$}
\end{minipage}%
}
\subfigure{
\begin{minipage}[t]{0.3\linewidth}
\centering
\includegraphics[width=\textwidth]{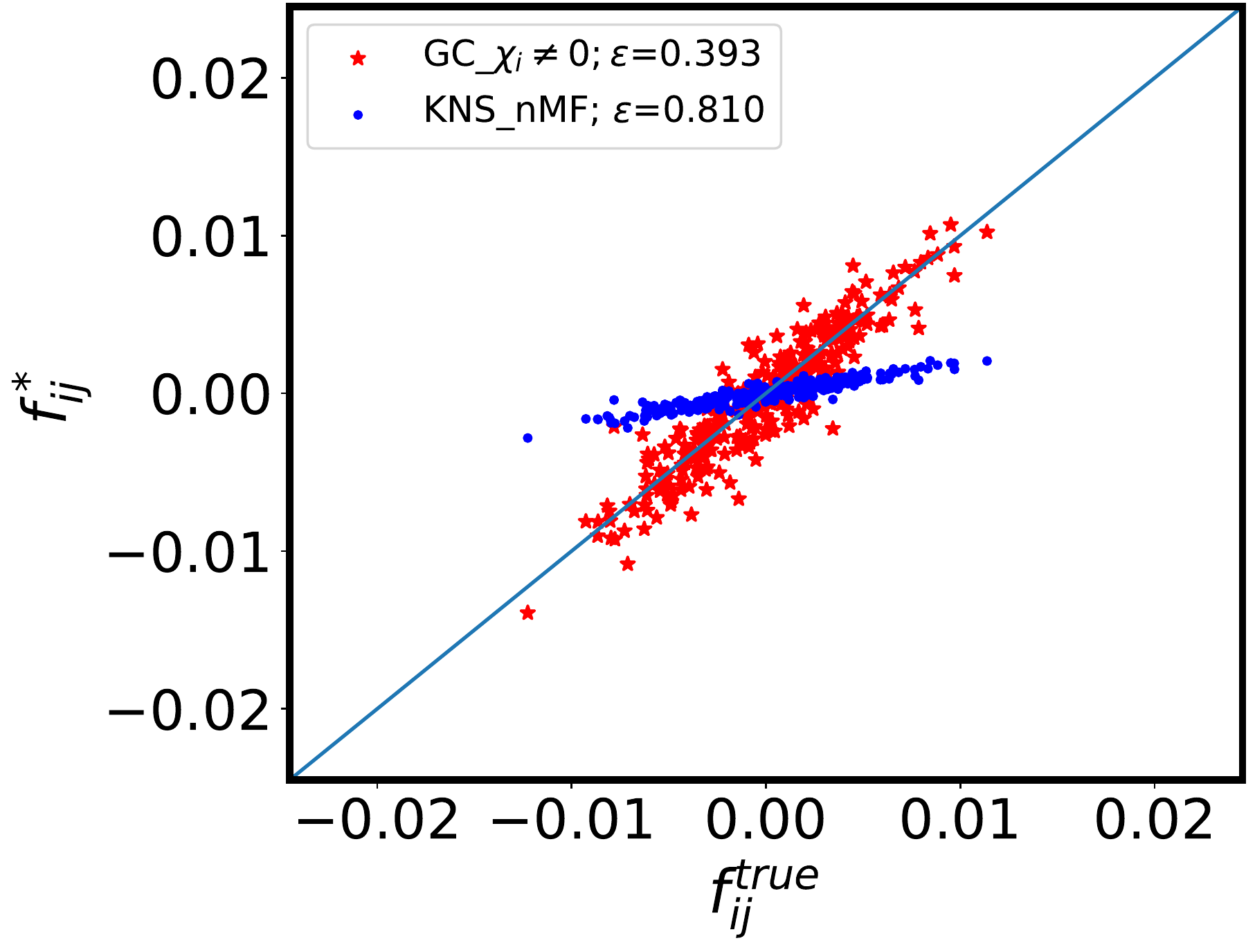}
\put(-101,-11){\small (e)~$\mu=0.25, r=0.5$}
\end{minipage}%
}
\subfigure{
\begin{minipage}[t]{0.3\linewidth}
\centering
\includegraphics[width=\textwidth]{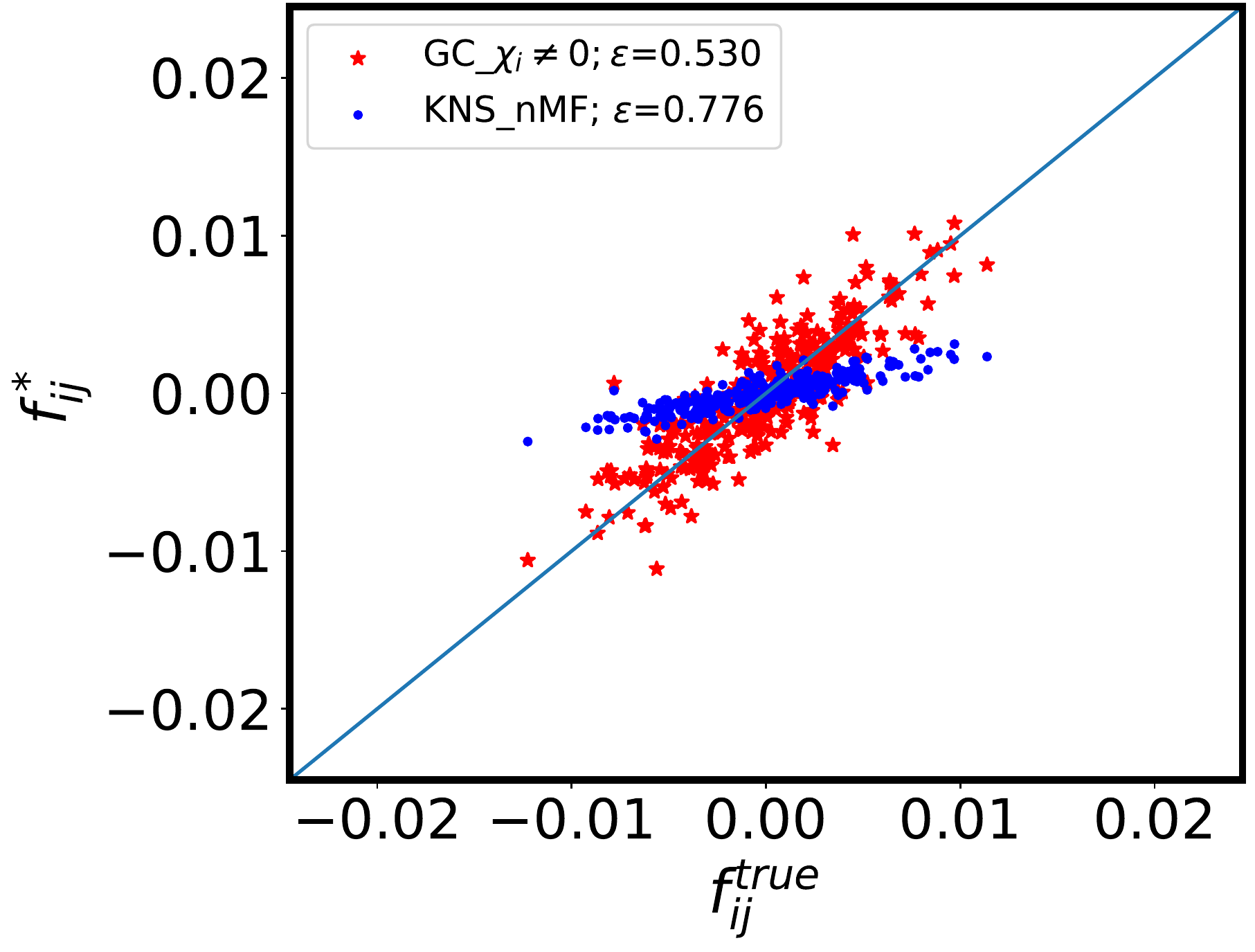}
\put(-101,-11){\small (f)~$\mu=0.25, r=0.8$}
\end{minipage}%
}\\
\subfigure{
\begin{minipage}[t]{0.3\linewidth}
\centering
\includegraphics[width=\textwidth]{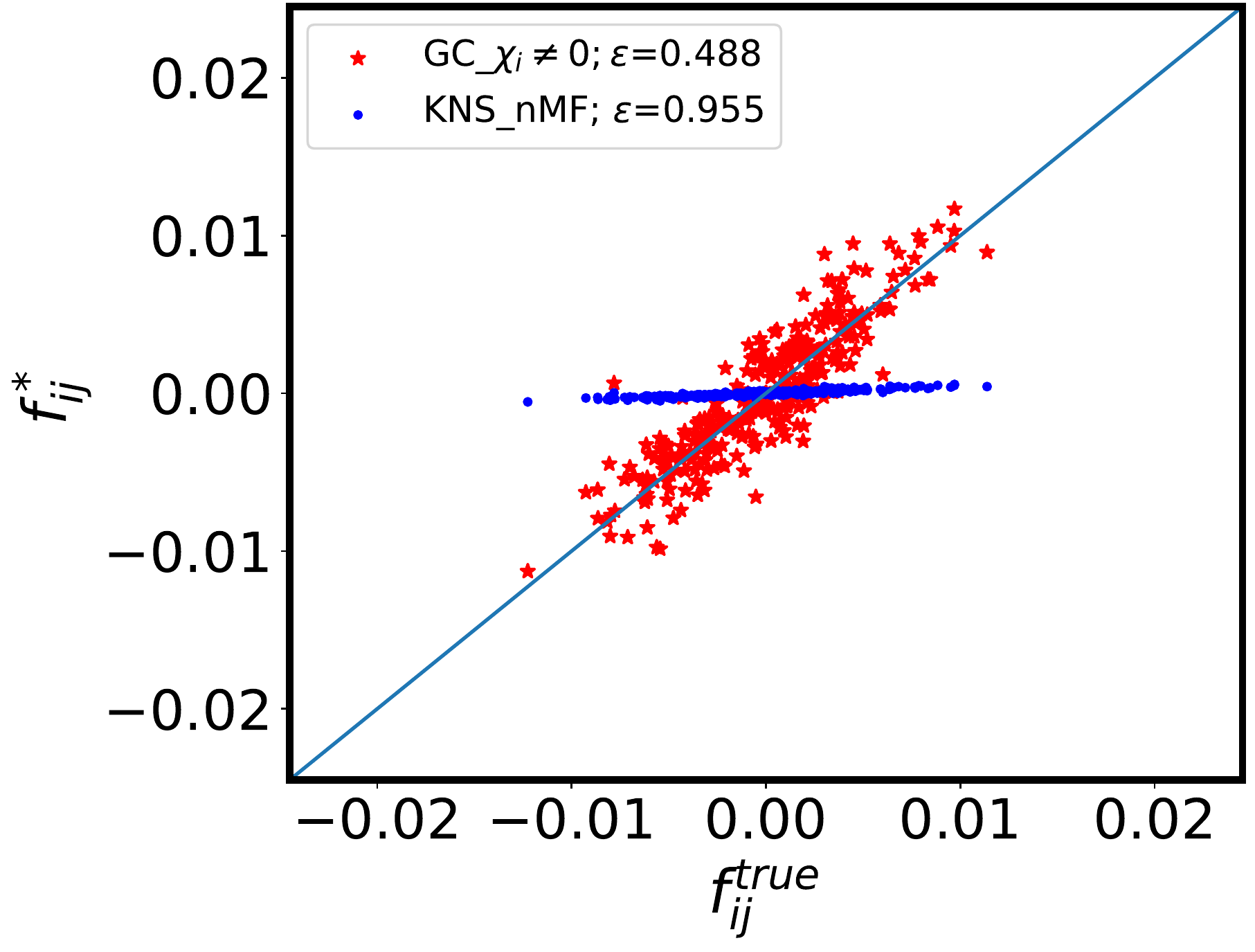}
\put(-101,-11){\small (g)~$\mu=0.5, r=0.2$}
\end{minipage}%
}
\subfigure{
\begin{minipage}[t]{0.3\linewidth}
\centering
\includegraphics[width=\textwidth]{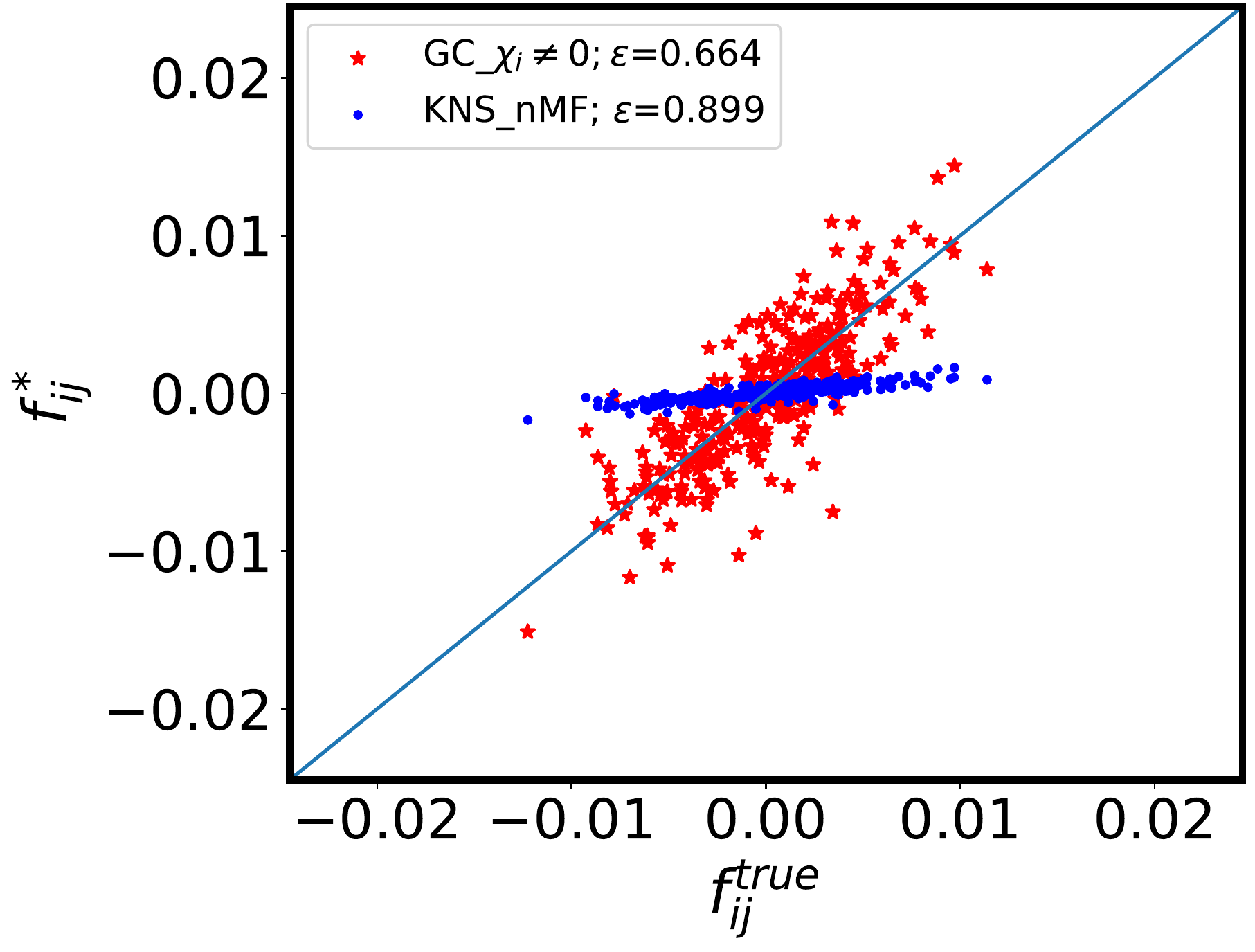}
\put(-101,-11){\small (h)~$\mu=0.5, r=0.5$}
\end{minipage}%
}
\subfigure{
\begin{minipage}[t]{0.3\linewidth}
\centering
\includegraphics[width=\textwidth]{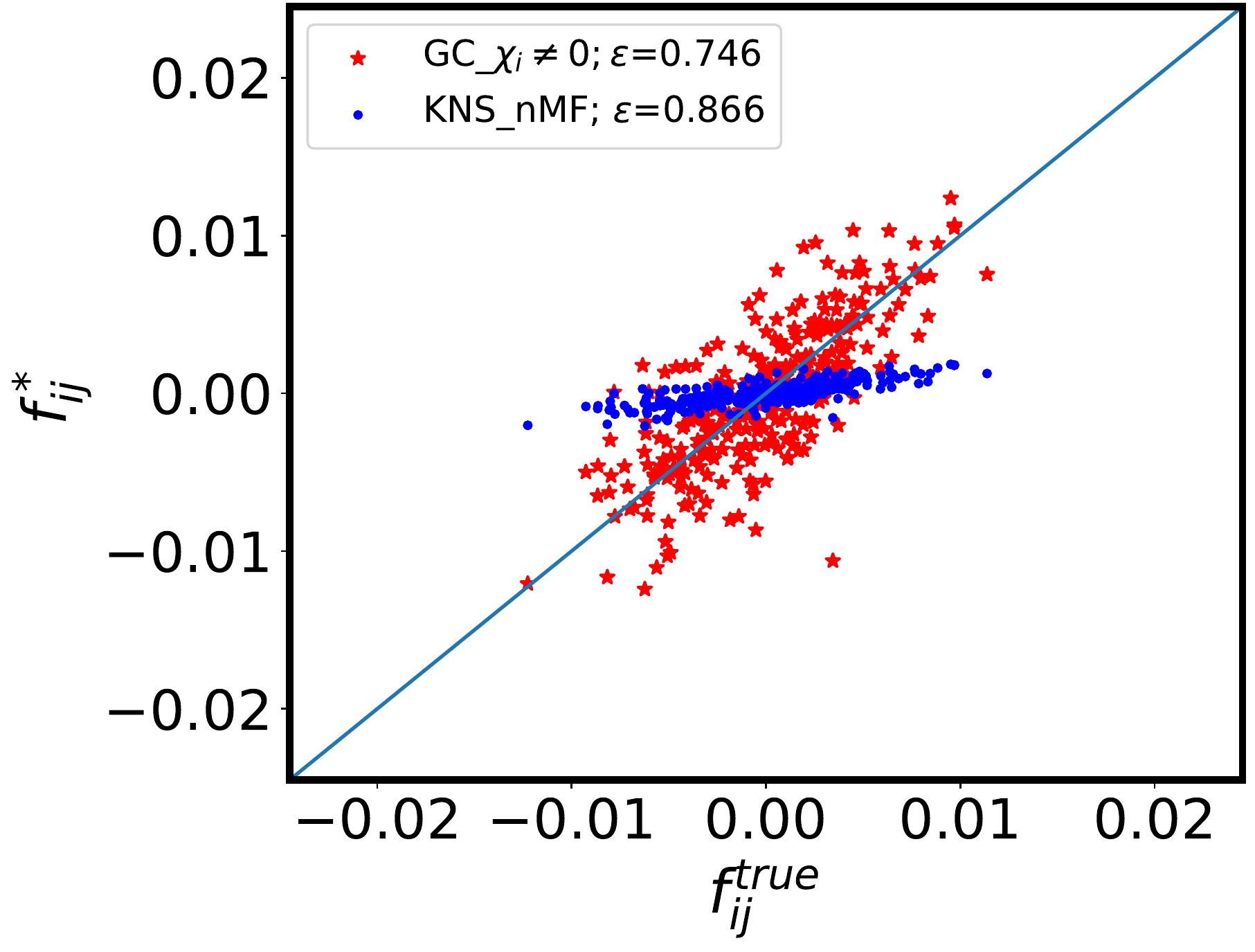}
\put(-101,-11){\small (i)~$\mu=0.5, r=0.8$}
\end{minipage}%
}
\caption{Scatter plots for testing and recovered $f_{ij}$s with mutation rate $\mu$ and recombination rate $r$. $r$ increases from left to right columns ($0.2, 0.5$ and $0.8$ respectively) while $\mu$ enlarge from top to bottom ($0.05, 0.25$ and $0.5$ respectively). The red stars for the Gaussian closed KNS $f_{ij}^* = \chi_{ij}\cdot (4\mu+ rc_{ij})/\left((1-\chi_i^2)(1-\chi_j^2)\right) $;  blue dots for original KNS $f_{ij}^* = rc_{ij} \cdot J_{ij}^{*,nMF}$.
Other parameters: $\sigma(\{f_{i}\})=0.05$, $\sigma(\{f_{ij}\})=0.004$,   cross-over rate $\rho=0.5$, number of loci $L=25$, carrying capacity $N=200$, number of generations $T=10,000$. Inference by Gaussian closed KNS works in much wider parameter range than original KNS. One realization of the fitness terms $f_{ij}$ and $f_i$ for each parameter value.}
\label{fig:scatter_diff_mu}
\end{figure*}

We start by taking a fixed fitness landscape
(same $f_{ij}$) and
systematically vary mutation and
recombination ($\mu$ and $r$).
Each sub-figure in large Fig.~\ref{fig:scatter_diff_mu}
shows scatter plots for the KNS
fitness inference formula  \eqref{eq:original_KNS_equation}
and the formula  \eqref{eq:fitness-inference-formula} based
on Gaussian closure
vs the model parameter $f_{ij}$ used to
generate the data.
These model parameters were independent Gaussian random variables
specified by their standard deviation $\sigma(\{f_i\})$ and $\sigma(\{f_{ij}\})$
as hyper-parameters.
The parameters $J_{ij}^*$ which enter
\eqref{eq:original_KNS_equation}
are inferred by naive mean-field (nMF).

The variations in  Fig.~ \ref{fig:scatter_diff_mu}
are such that each column has the same recombination
rate in the order low-medium-high from left to right,
and each row has the same mutation rate
in the order low-medium-high from top to bottom.
In the top row both inference formulae
work well, particularly for high recombination rate at the top right.
In the middle and bottom rows the KNS formula does not work while
the formula based
on Gaussian closure
still performs well, and in particular does
not have systematic errors.

For comparison in more
extensive parameter ranges we have
quantified inference performance by
normalized root of mean square error
\begin{equation}
    \epsilon = \sqrt{\frac{\sum_{ij}\left(f^*_{ij}-f_{ij}\right)^2}{\sum_{ij}f_{ij}^2}}
    \label{eq:epsilon}
\end{equation}
We note that this reduces all the information in
the scatter plots in Fig.~\ref{fig:scatter_diff_mu}
to one single number.
Although we have not observed
such behaviour, it
is conceivable that inference could be very accurate
for most pairs ($i,j$) such that $\epsilon$
is small, but
still have large errors for some few
pairs. An overall value $\epsilon$
much less than one hence does not
guarantee that fitness inference
is accurate for all pairs.
On the other hand, a large mean square error
could correspond to either systematic
or random errors in the scatter plots. 
Both behaviours we have observed.

\begin{figure}[!ht]
            \centering
            \includegraphics[width=0.9\textwidth]{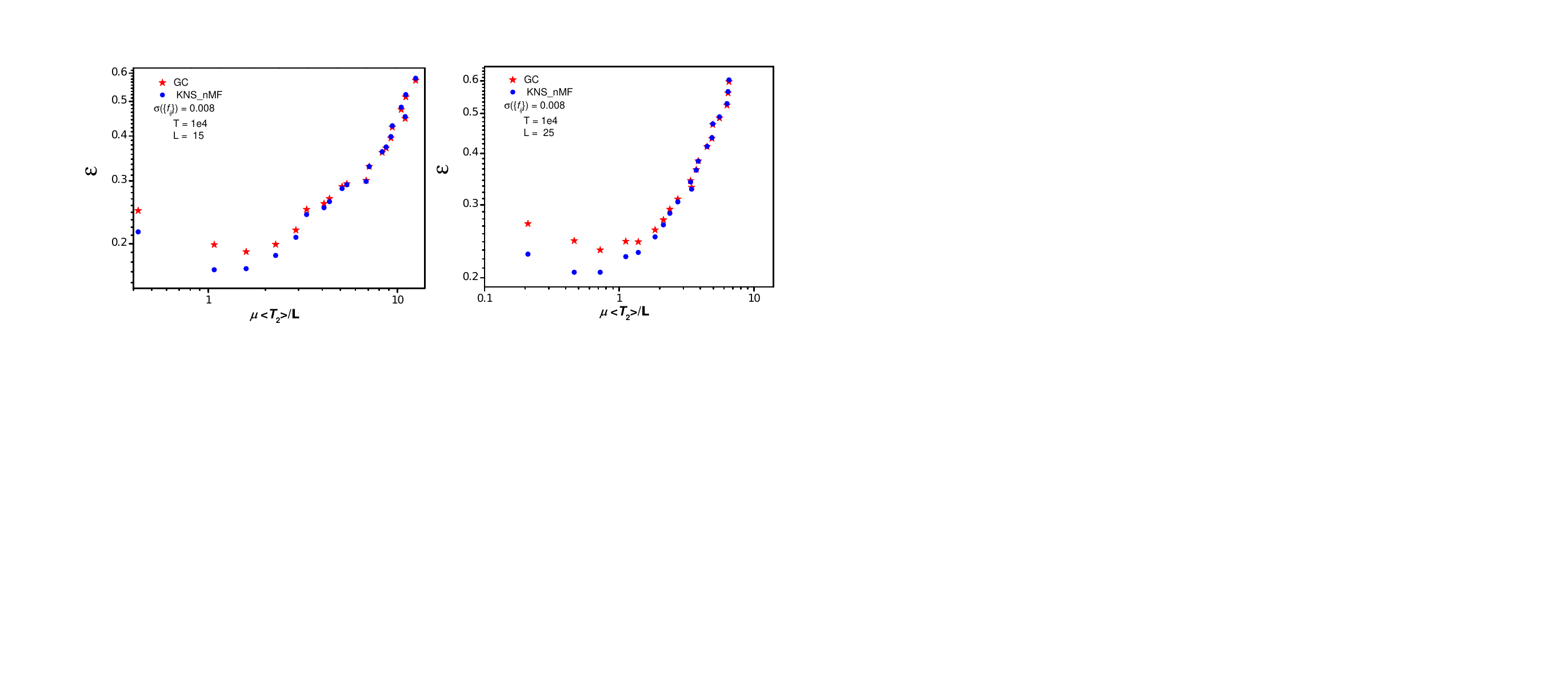}
            \caption{Epistasis reconstruction error $\epsilon$ versus $\mu\langle T_2\rangle/L$. Red stars for $f_{ij}^*=\chi_{ij}\cdot(4\mu+rc_{ij})/((1-\chi_i^2)(1-\chi_j^2))$ while blue dots for  $f_{ij}^*=J_{ij}^{*,nMF}\cdot(4\mu+rc_{ij})$. Epistasis $f_{ij}$ are inferred best with $\mu=0.05$.
            The other parameter values:  $\sigma(f_i) = 0.05$, $\sigma(f_{ij}) = 0.008$, carrying capacity $N = 200$, out-crossing rate $r=0.5$, cross-over rate $\rho= 0.5$, number of loci $L = 15$, generations $T = 10,000$. 10 realizations of the fitness terms $f_{ij}$ and $f_i$ for each parameter value. }
            \label{fig:epsilon_mu_T2}
        \end{figure}

Anticipating a discussion which we 
give 
in Appendix~\ref{a:FFPopSim}
we chose to visualize the dependence of
$\epsilon$ on variation of $\mu$
by incorporating the coalescence
time $\left<T_2\right>$, previously 
used in theoretical discussions
of problems of the kind studied here
\cite{Neher2013,Held2019}.
Fig.~\ref{fig:epsilon_mu_T2}
shows that is, at least for low
epistatic fitness, a tendency 
reconstruction error to grow with
$\mu$. For the tests that have been carried out
it also appears that the threshold 
between the phase where fitness inference
is possible takes place when there is
about one mutation per coalescence time
and number of pairs of loci.

\begin{figure*}[!ht]
\centering
\includegraphics[width=0.45\textwidth]{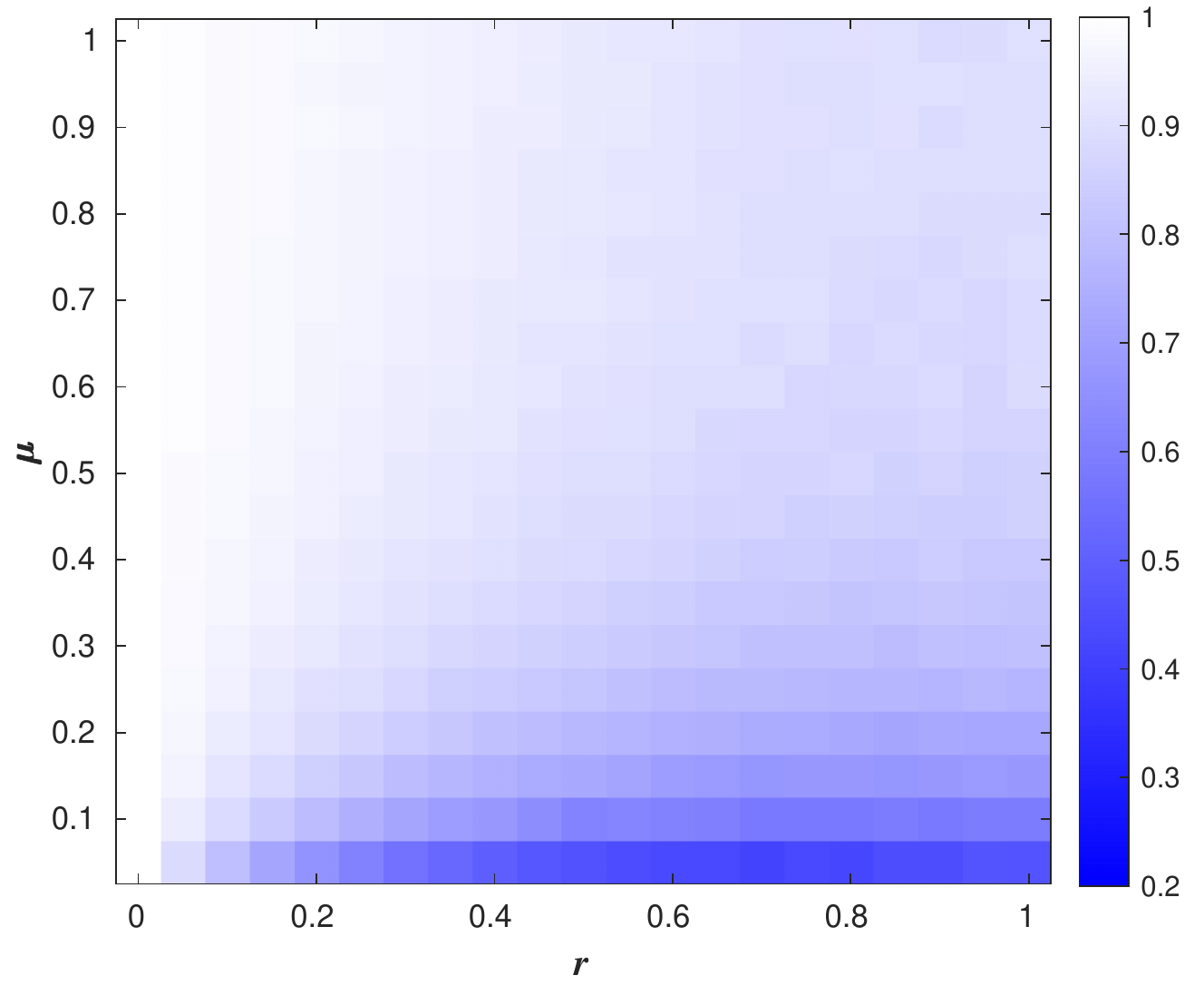}
\includegraphics[width=0.45\textwidth]{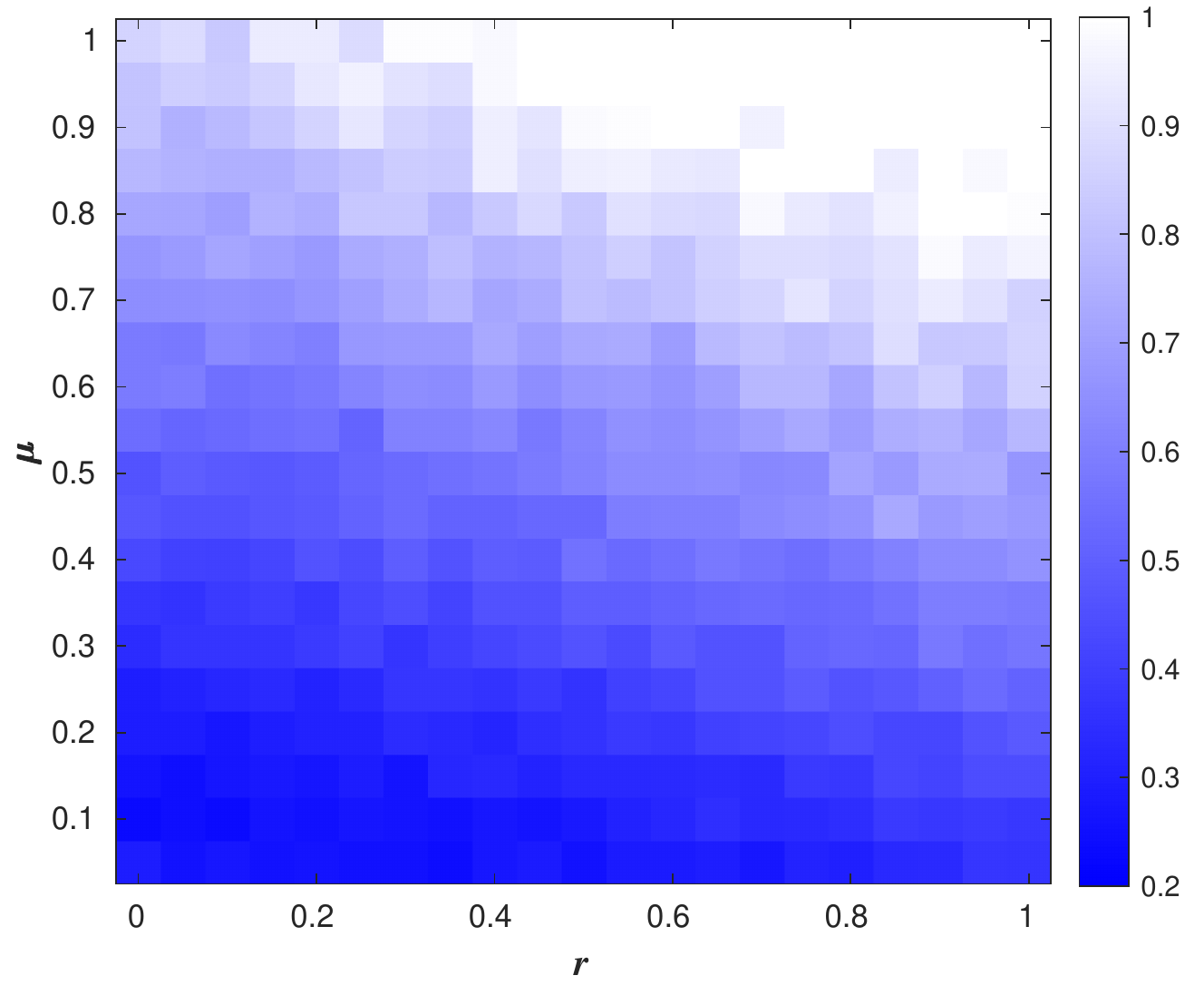}
\caption{Phase diagram for mutation rate $\mu$ versus recombination rate $r$. The color is encoded by the reconstruction error $\epsilon$ given in eq. \eqref{eq:epsilon}. Left: KNS theory $f_{ij} =J_{ij}^{*,nMF} \cdot rc_{ij} $. Right: Gaussian closed KNS theory $f_{ij} = \chi_{ij} \cdot (4\mu + rc_{ij})/\left((1-\chi_i^2)(1-\chi_j^2)\right)$. Parameters:$\sigma(\{f_{i}\})=0.05$, $\sigma(\{f_{ij}\})=0.004$,   cross-over rate $\rho=0.5$, number of loci $L=25$, carrying capacity $N=200$, generations $T=10,000$. One realization of the fitness terms $f_{ij}$ and $f_i$ for each parameter value.}
\label{fig:phase_plots_mu_r_T10000}
\end{figure*}        
Returning now to the mapping out
of regions where parameter inference is
possible or not possible, we point to phase diagrams
of $\epsilon$ shown in Fig. \ref{fig:phase_plots_mu_r_T10000},
for respectively KNS
formula with nMF and the formula from Gaussian closure.
Number of generations in simulations is set as
$T = 10,000$ and kept as a constant for all combinations of parameters.
As in the scatter plots we observe
large differences as
to two epistatic fitness inference formulae.
In short, for linear structure of genomes,
the KNS formula \eqref{eq:original_KNS_equation} works only for low mutation rate  and high recombination rate (Fig. \ref{fig:scatter_diff_mu}c).
The new formula \eqref{eq:fitness-inference-formula} from Gaussian closure
instead works for
a much larger region with weak fitness. The standard deviation of epistatic fitness $\sigma\left(\{f_{ij}\}\right)=0.004$ in Fig. \ref{fig:phase_plots_mu_r_T10000}.
We comment on the reasons
for this effect in Section~\ref{sec:discussion}.
For stronger mutation rate and larger recombination rate (data not shown) the
root mean square error ($\epsilon$s)
of inference based on the Gaussian closure
formula increases,
\textit{i.e.} in that range this formula does not work either. Specifically, the KNS formula \eqref{eq:original_KNS_equation} has severe systematic error while the formula \eqref{eq:fitness-inference-formula} with Gaussian closure performs worse with heavier noise.

\subsection{Fitness variations vs recombination rate}
\label{sec:fitness-vs-recombination}
\begin{figure*}[!ht]
\centering
\subfigure{
\begin{minipage}[t]{0.31\linewidth}
\centering
\includegraphics[width=\textwidth]{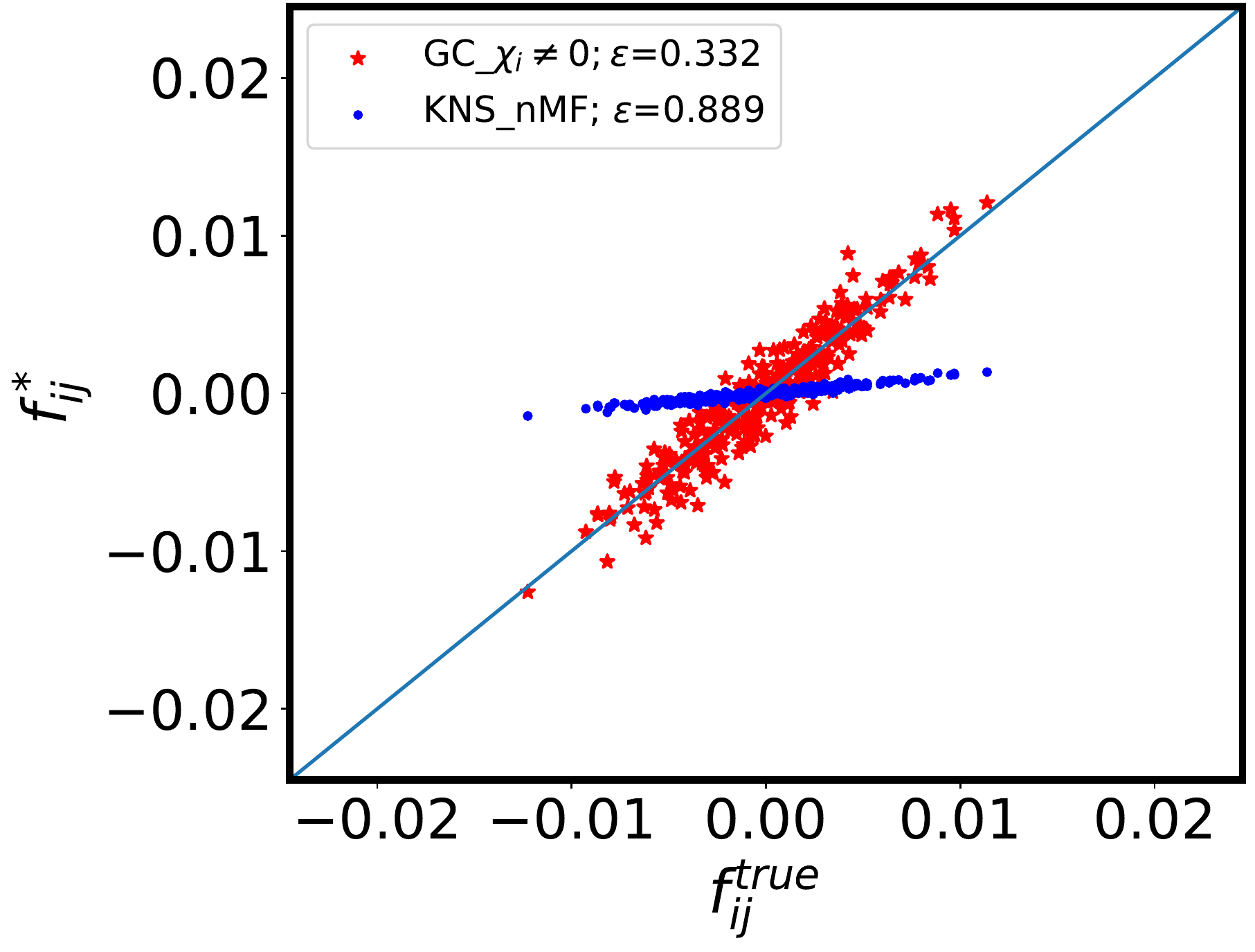}
\put(-101,-11){\small (a)~$\sigma=0.004, r=0.2$}
\end{minipage}%
}
\subfigure{
\begin{minipage}[t]{0.31\linewidth}
\centering
\includegraphics[width=\textwidth]{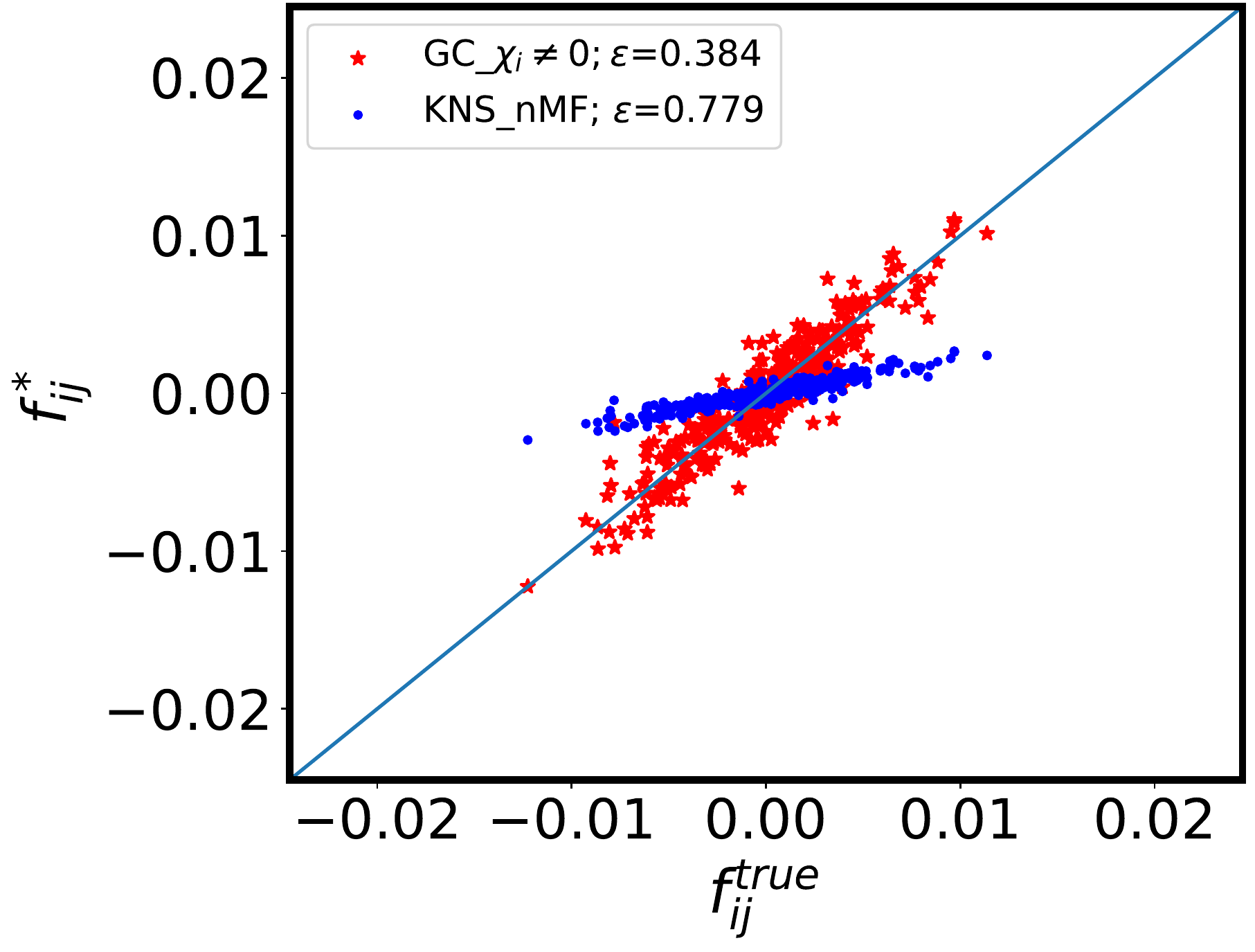}
\put(-101,-11){\small (b)~$\sigma=0.004, r=0.5$}
\end{minipage}%
}
\subfigure{
\begin{minipage}[t]{0.31\linewidth}
\centering
\includegraphics[width=\textwidth]{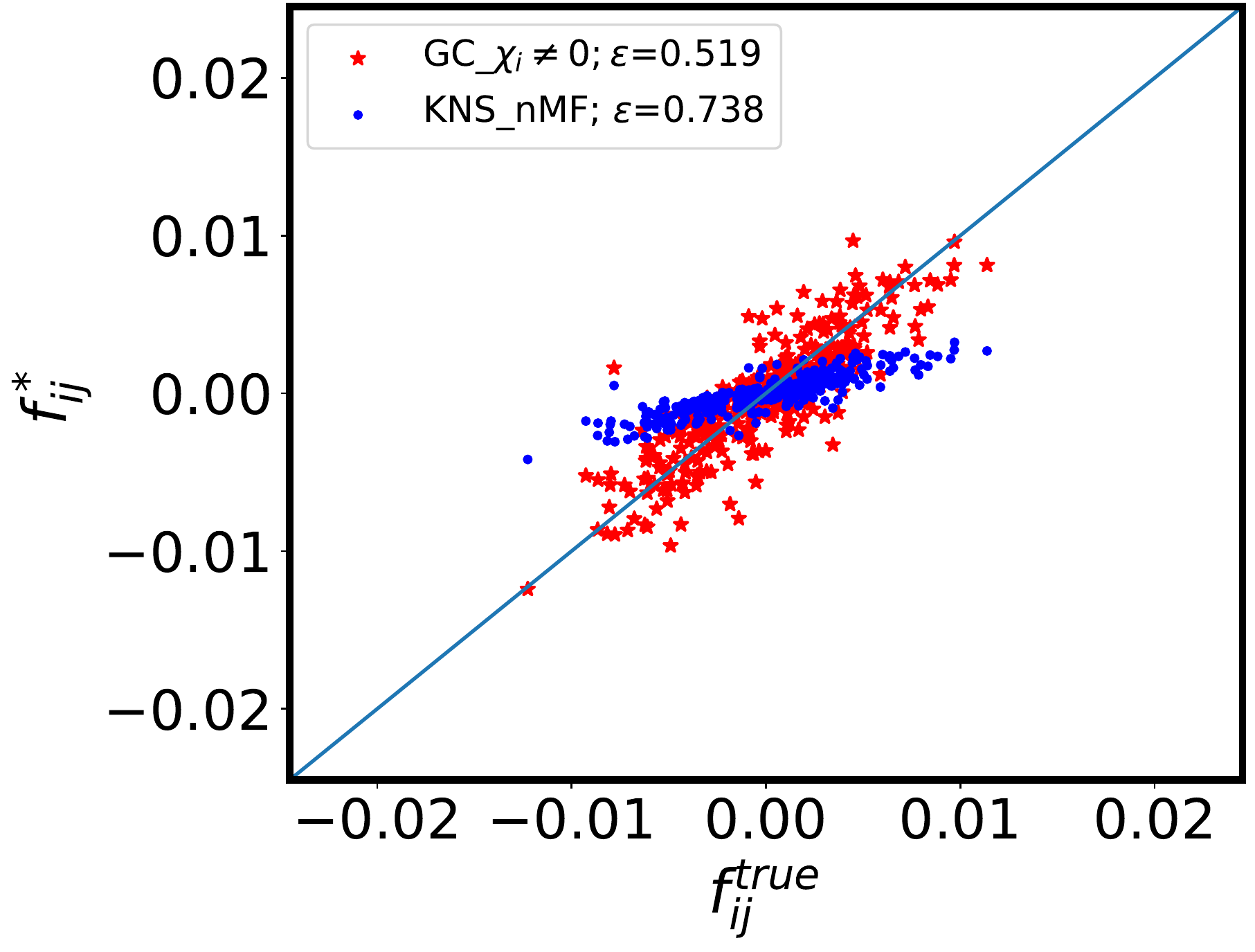}
\put(-101,-11){\small (c)~$\sigma=0.004, r=0.8$}
\end{minipage}%
}\\
\subfigure{
\begin{minipage}[t]{0.31\linewidth}
\centering
\includegraphics[width=\textwidth]{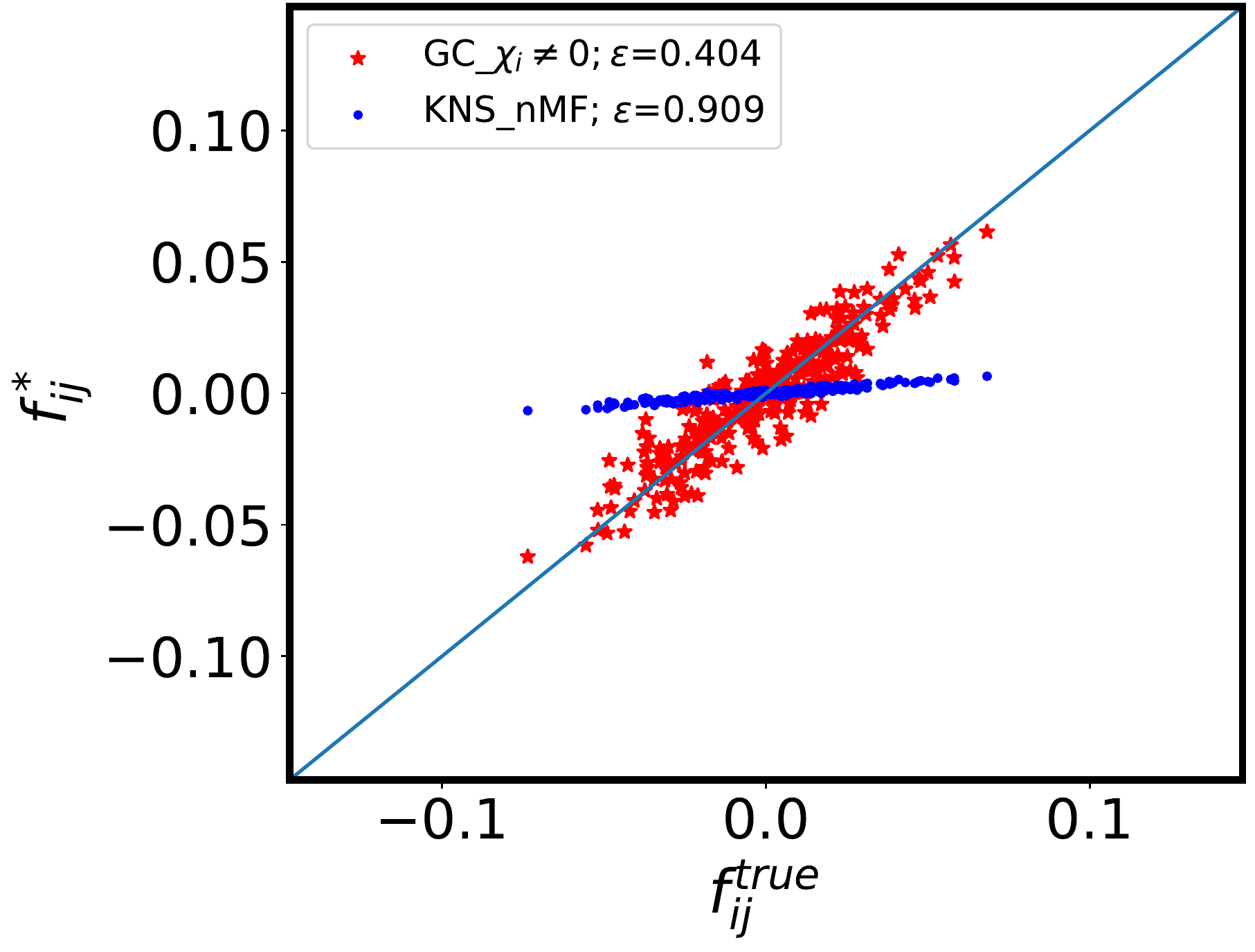}
\put(-101,-11){\small (d)~$\sigma=0.024, r=0.2$}
\end{minipage}%
}
\subfigure{
\begin{minipage}[t]{0.31\linewidth}
\centering
\includegraphics[width=\textwidth]{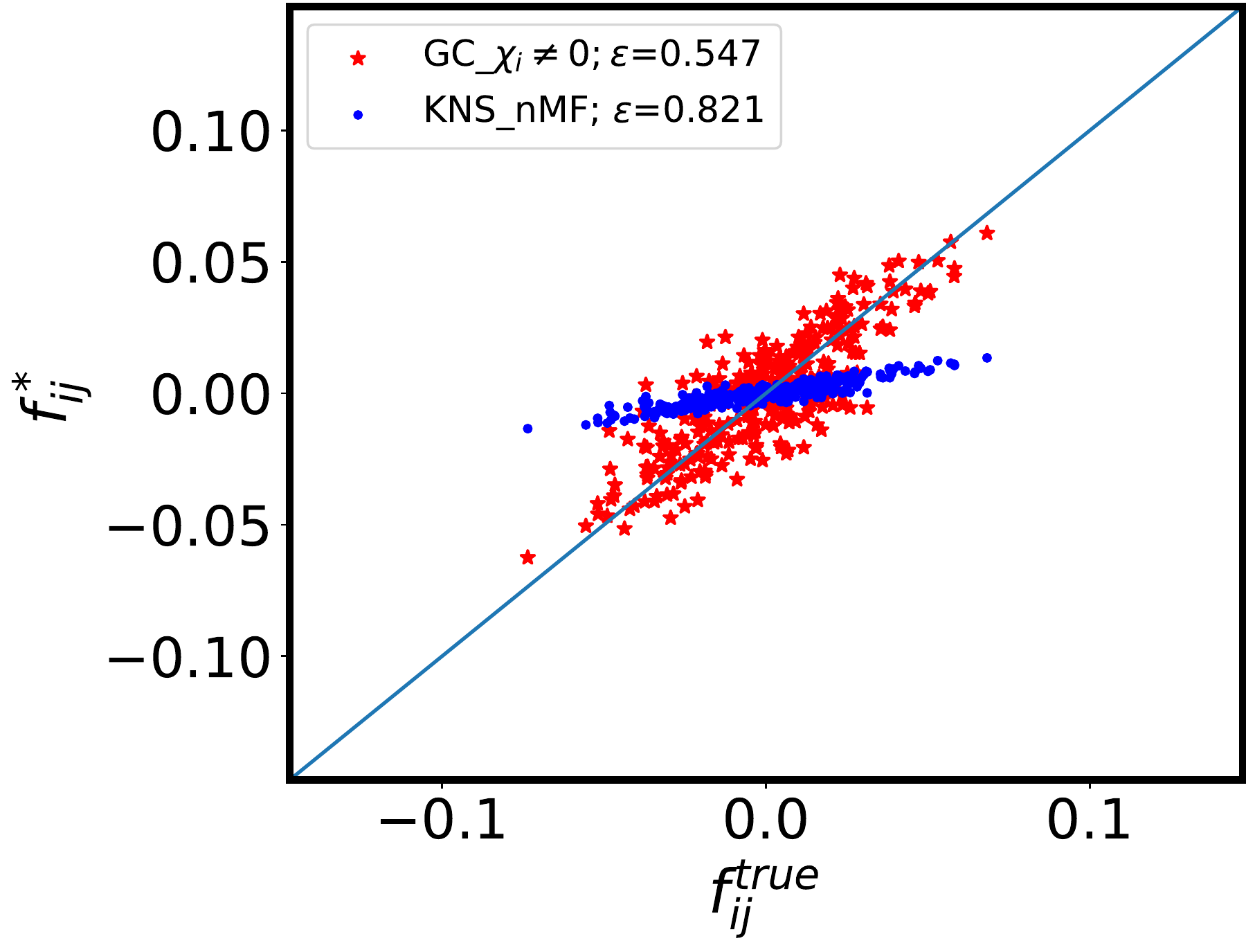}
\put(-101,-11){\small (e)~$\sigma=0.024, r=0.5$}
\end{minipage}%
}
\subfigure{
\begin{minipage}[t]{0.31\linewidth}
\centering
\includegraphics[width=\textwidth]{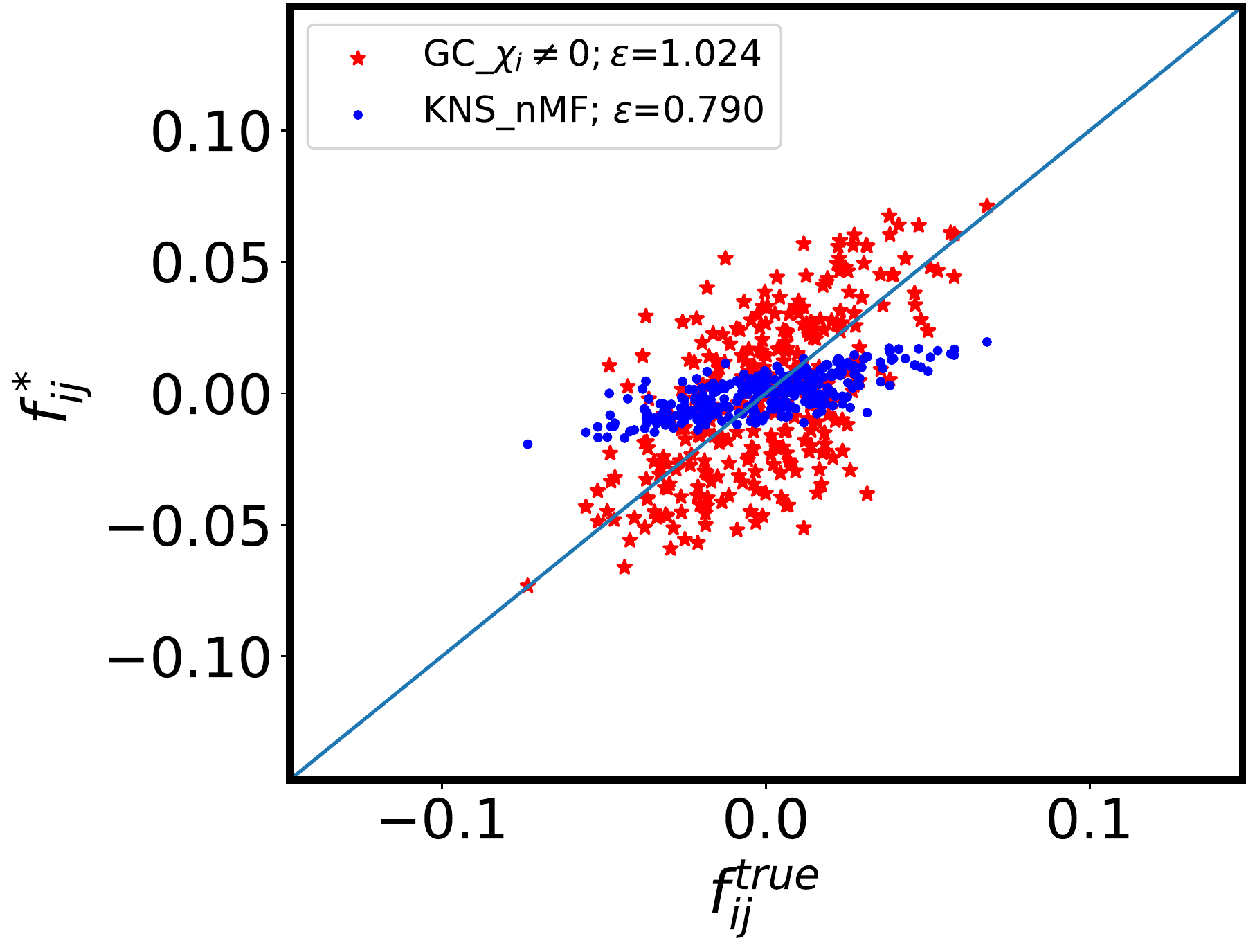}
\put(-101,-11){\small (f)~$\sigma=0.024, r=0.8$}
\end{minipage}%
}\\
\subfigure{
\begin{minipage}[t]{0.31\linewidth}
\centering
\includegraphics[width=\textwidth]{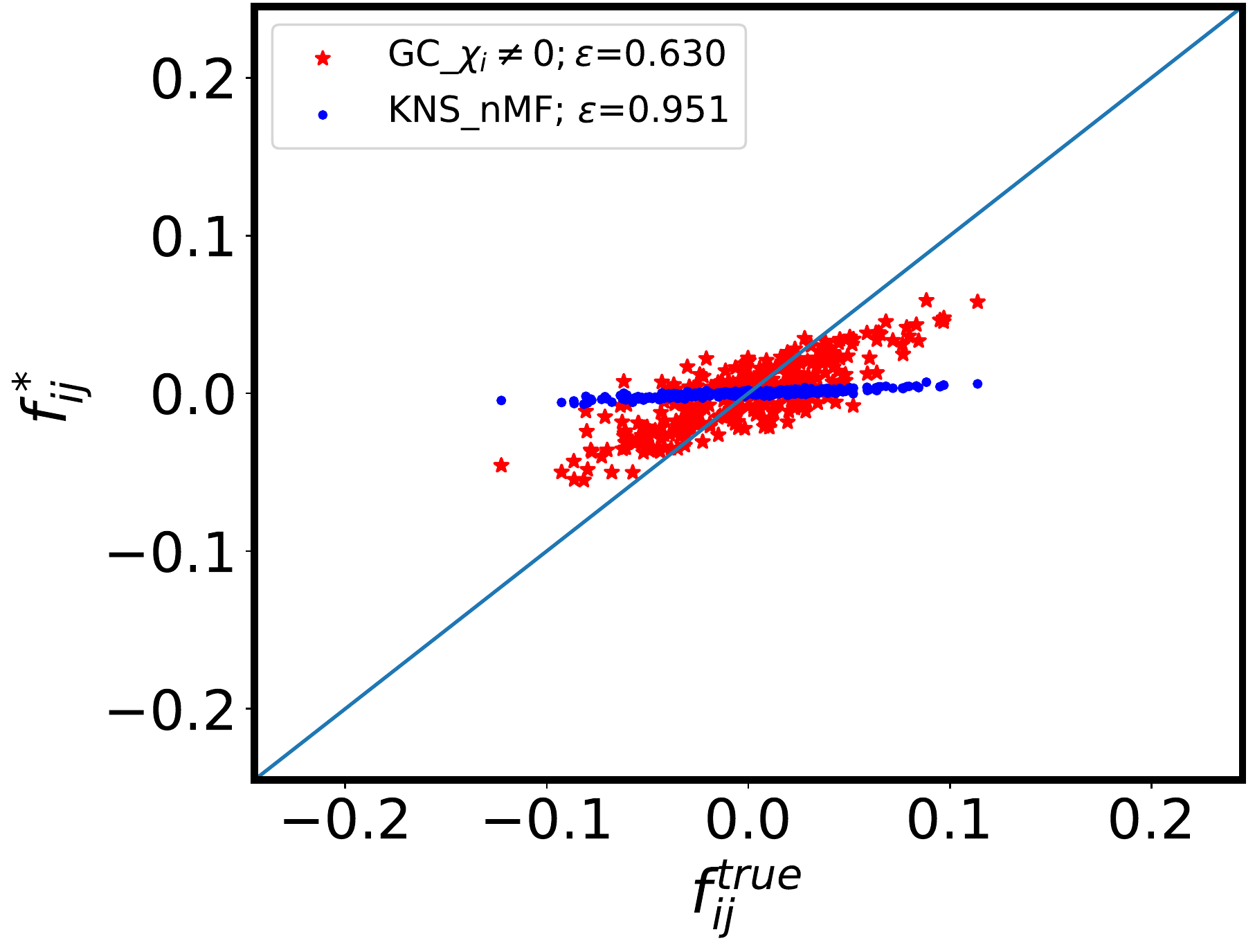}
\put(-101,-11){\small (g)~$\sigma=0.04, r=0.2$}
\end{minipage}%
}
\subfigure{
\begin{minipage}[t]{0.31\linewidth}
\centering
\includegraphics[width=\textwidth]{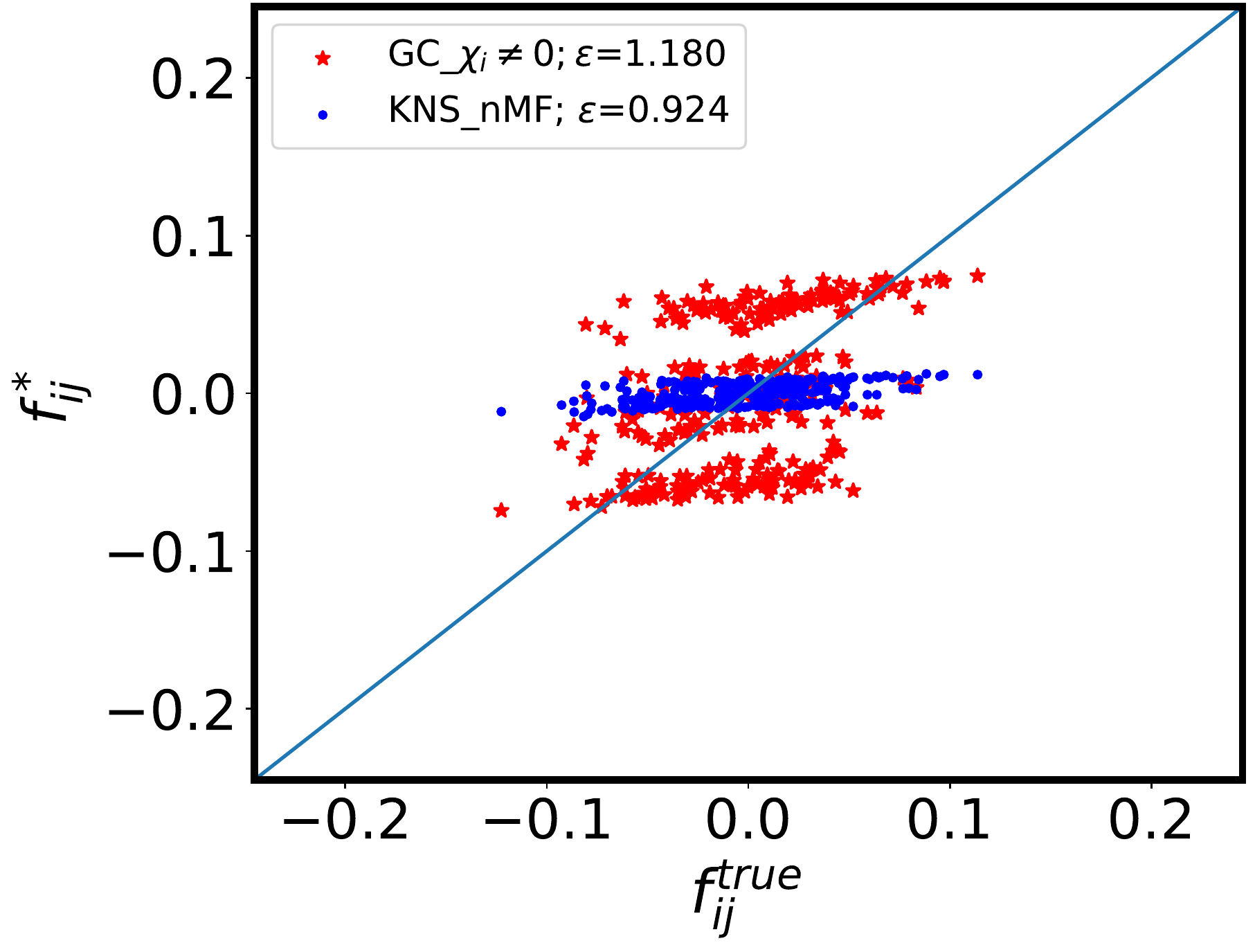}
\put(-101,-11){\small (h)~$\sigma=0.04, r=0.5$}
\end{minipage}%
}
\subfigure{
\begin{minipage}[t]{0.31\linewidth}
\centering
\includegraphics[width=\textwidth]{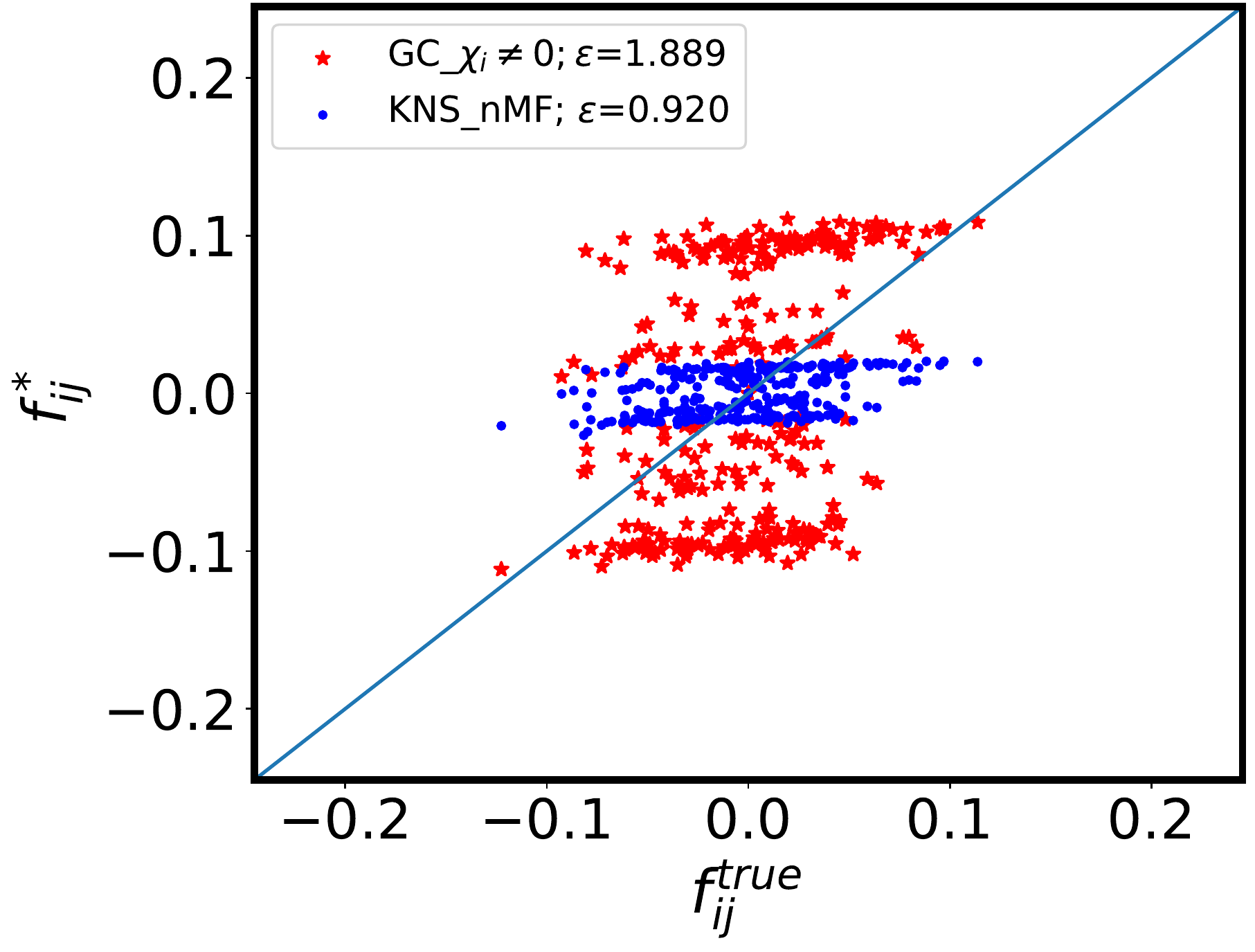}
\put(-101,-11){\small (i)~$\sigma=0.04, r=0.8$}
\end{minipage}%
}
\caption{Scatter plots for testing and reconstructed $f_{ij}$s. The standard deviation $\sigma(\{f_{ij}\}^{true})$ increases from top to bottom rows ($0.004, 0.024$ and $0.04$ respectively) and recombination rate $r$ enlarges in columns from left to right ($0.2, 0.5$ and $0.8$ respectively). Red stars for $f_{ij}^* = \chi_{ij} \cdot (4\mu + rc_{ij})/((1-\chi_i^2)(1-\chi_j^2))$ and blue dots for $f_{ij}^*=J_{ij}^{*,nMF} \cdot rc_{ij}$. Both inference formulae do not work for large $\sigma$ and high $r$, where strong correlations emerge between loci that drive the system out of the QLE phase \cite{Dichio2020,DZA}, as shown in (g), (h) and (i).
The other parameter values: standard deviation $\sigma(\{f_i\})=0.05$, mutation rate $\mu=0.2$,  cross-over rate $\rho=0.5$, number of loci $L=25$, carrying capacity $N=200$, generations $T=10,000$.  One realization of the fitness terms $f_{ij}$ and $f_i$ for each parameter value.}
\label{fig:Scatter-plots-diff-sigma}
\end{figure*}

We continue by varying recombination
$r$ and the dispersion
in the fitness landscape
($f_{ij}$ drawn from Gaussian distributions
with different hyper-parameters $\sigma(\{f_{ij}\})$).
Each sub-figure in Fig.~\ref{fig:Scatter-plots-diff-sigma}
shows scatter plots for the
two epistatic fitness inference formulae
for the model parameter $\sigma\left(\{f_{ij}\}\right)$ vs recombination rate $r$.
The order in the Fig. \ref{fig:Scatter-plots-diff-sigma}
is increasing recombination
rate $r$ in columns from left to right,
and increasing $\sigma(\{f_{ij}\})$
in rows from top to bottom.
Here, the mutation rate $\mu=0.2$ and the other parameters are the same with those tested in Fig. \ref{fig:scatter_diff_mu}.

\begin{figure*}[!ht]
\centering
\includegraphics[width=0.45\textwidth]{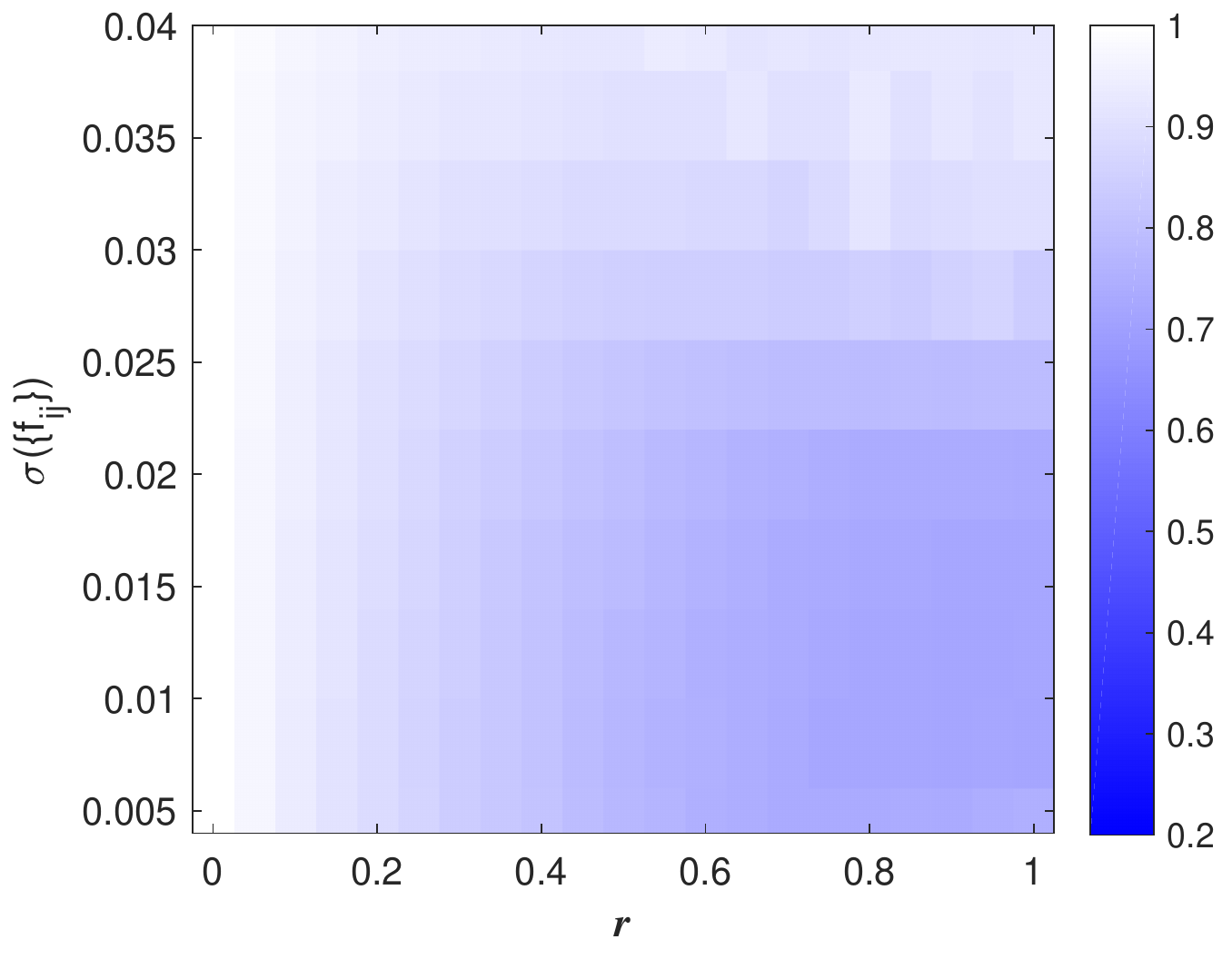}
\includegraphics[width=0.45\textwidth]{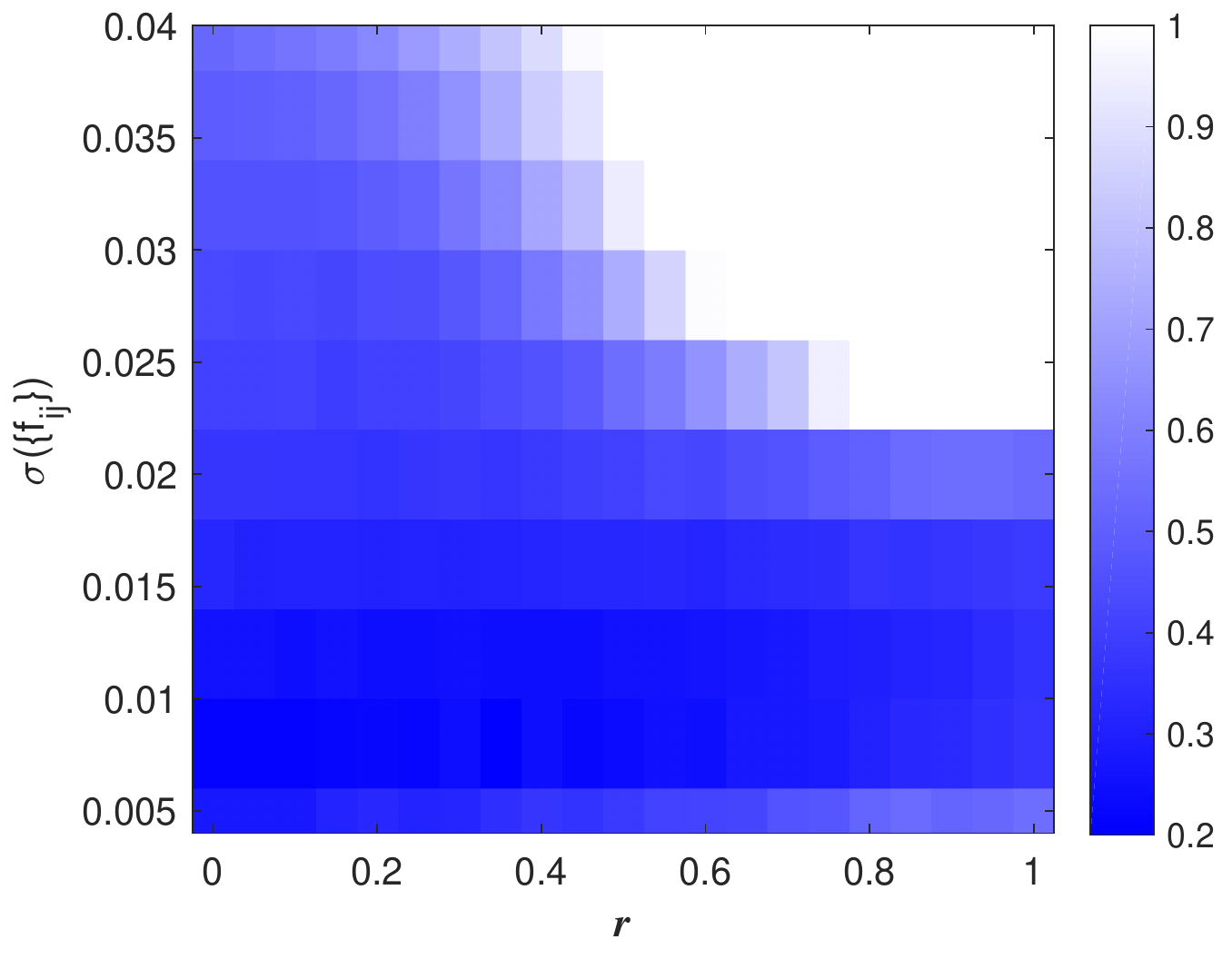}
\caption{Phase diagram for the standard deviation $\sigma(\{f_{ij}\})$ versus recombination rate $r$. Left: KNS theory $f_{ij}^* = J_{ij}^{nMF} \cdot rc_{ij}$. Right: Gaussian closed KNS theory $f_{ij}^* = \chi_{ij} \cdot (4\mu + rc_{ij})/((1-\chi_i^2)(1-\chi_j^2))$. Parameters: mutation rate $\mu=0.2$,  cross-over rate $\rho=0.5$, number of loci $L=25$, carrying capacity $N=200$, generations $T=10,000$. One realization of the fitness terms $f_{ij}$ and $f_i$ for each parameter value.}
\label{fig:figs-sigma-r}
\end{figure*}

Overall, the KNS formula \eqref{eq:original_KNS_equation}
does not work for any of the parameter
values shown in
Fig.~\ref{fig:Scatter-plots-diff-sigma} with mutation rate $\mu=0.2$.

This
either because of systematic errors
as in top row (low fitness dispersion)
and left column (low recombination),
or due to the emergence of strong correlation between loci that drive the evolution out of the QLE regime \cite{Dichio2020,DZA},
as in bottom right corner Fig.~\ref{fig:Scatter-plots-diff-sigma}h and Fig.~\ref{fig:Scatter-plots-diff-sigma}i.
The Gaussian closure formula \eqref{eq:fitness-inference-formula} in contrast works well for low recombination or low fitness dispersion, or both but fails as well for sufficiently high recombination and fitness strength.

As above we have quantified
inference performance in larger
parameter ranges by the root of mean square error $\epsilon$.
The phase diagrams in Fig.~\ref{fig:figs-sigma-r}
show again that the Gaussian closure formula
works except when $r$ and $\sigma(\{f_{ij}\})$
are both large, while the KNS
formula does not work in any range with mutation rate $\mu=0.2 $.

\section{Discussion}
\label{sec:discussion}
In this paper we have pursued the
investigations started by Kimura
in 1965~\cite{Kimura1965}
on how epistatic contributions
to fitness is reflected in the
distribution over genotypes in a population.
Our perspective is that of fitness inference:
we assume that the distribution is observable
from many whole-genome sequences of
an organism and ask what we can
learn about synergistic effects on fitness
from concurrent allele variations at
different loci, \textit{i.e.} about epistasis.
Our benchmark has been the generalization
of the Kimura theory by Neher and
Shraiman to a phase of genome-scale
Quasi-Linkage Equilibrium (QLE)~\cite{NeherShraiman2009,NeherShraiman2011,Gao2019}.
In recent work we showed in numerical
testing that a central formula
describing the QLE phase allows to retrieve
epistatic contributions to fitness in
the limit of high recombination~\cite{Zeng2020}.

Here we have extended these considerations
to the regime where mutation can be a stronger (faster)
process than recombination. We have done this on one hand
by generalizing the derivation from the assumptions of a QLE
state in~\cite{NeherShraiman2011}, and on the other
by following the analogy to physical
kinetics and approximations of the Boltzmann equation, recently developed by three of us~\cite{Mauri-2019,Mauri-2020}.
In the second approach we
consider the evolution equations
for single-locus and two-loci frequencies
in a population evolving under selection, mutation,
genetic drift and recombination,
and then close those equations
by setting higher-order cumulants to zero
(Gaussian closure).

We hence do not need to make
explicit assumptions
on the functional form of the distribution
over genotypes in a population, only that
it is possible to treat it as a Gaussian
for the purpose of evaluating higher moments.
Methods of a similar type were earlier developed
for a more macroscopic level approach that does
not assume access to individual genotypes,
and therefore also cannot be tested on that level
\cite{Nourmohammad2013}.

The phase diagrams in
Fig.~\ref{fig:phase_plots_mu_r_T10000}
and Fig.~\ref{fig:figs-sigma-r}
in Section~\ref{sec:simulation}
show that the new inference formula
dramatically outperforms
previous one, with the
exception of a region at strong fitness and high
recombination rate.
Since in many biological systems mutations can
be at least as strong as recombination, we have
extended the range of epistasis inference methods
considerably. This is the main result of the current work.
We have also performed preliminary investigations 
of a possible dependence on the phase boundary on
coalescence time, a quantity which roughly
measures the time to a common ancestor for the
whole population. In earlier theoretical work
based either on the infinitesimal model
of genetics or on more macroscopic considerations,
it has been predicted that the transition 
occurs at about one mutation per coalescence time
per locus pair \cite{Neher2013,Held2019}.
To the extent that we have been able to test 
this hypothesis we find that it holds also
for our procedure where inference 
is assessed for each epistatic pair and
its associated epistatic fitness parameter 
separately, see Fig.~\ref{fig:epsilon_mu_T2}.

A theoretical advantage of the
extension presented here follows from the
fact that the previous inference formula
\eqref{eq:original_KNS_equation}
is obtained by perturbation
in the inverse of recombination rate
\textit{i.e.} eq.~(23) and Appendix B
in~\cite{NeherShraiman2011}.
In a finite population this derivation requires
that mutations be so much
weaker compared to recombination that they
can be neglected for quantitative properties
in QLE, while still being non-zero.
The latter restriction is necessary
as otherwise the fittest genotype will
eventually take over the population,
and the QLE phase will only be a long-lived
transient~\cite{Gao2019,Zeng2020}.
One consequence of a low mutation rate is
that any imbalance in total epistatic fitness will lead to almost fixated alleles.
In the QLE phase where eq. \eqref{eq:original_KNS_equation}
can be used quantitatively, the first order moments
$\chi_i$ are therefore typically different than zero.
Inference formula \eqref{eq:fitness-inference-formula}
is on the other hand obtained
by expanding the equations of Gaussian closure
under conditions appropriate for high mutation rate,
and $\chi_i$ are not necessarily to be zero neither.
A further assumption to arrive at
\eqref{eq:fitness-inference-formula} is that
epistatic fitness variations are not too strong,
qualitatively $L\sigma(\{f_{ij}\}) < 1$.
Moreover, the additive fitness should be sufficiently weak as well to make sure the population is strictly mono-clonal which is one of the assumptions of the Gaussian closure \cite{Mauri-2019, Mauri-2020}.
Data shown in bottom row of
Fig.~\ref{fig:Scatter-plots-diff-sigma}
(sub-figures (g), (h) and (i))
have $L\sigma(\{f_{ij}\}) \approx 1$.

In conclusion, we have presented
an extension of the classic Kimura-Neher-Shraiman
theory which allows to reliably infer
epistatic contributions to fitness.
In separate work three of us recently applied
the method to more than 50,000
full-length genomes of the SARS-CoV-2 virus~\cite{Zeng-COVID19},
and were able to to predict new epistatic interactions
between eight viral genes, many involving
ORF3a, a protein implicated in severe
manifestations of COVID19 disease.
Methodological development of epistasis analysis using DCA, as we have
discussed here, may hence also
have practical applications of some impact
on society.

\section*{Acknowledgements}
We are grateful to Guilhem Semerjian for valuable
input and suggestions. We also acknowledge
constructive criticism of an anonymous referees, which
allowed us to significantly improve our work.
The work of HLZ was sponsored by National Natural Science Foundation of China (11705097), Natural Science Foundation of Jiangsu Province (BK20170895), Jiangsu Government Scholarship for Overseas Studies of 2018.
The work of EM was supported by ICFP fellowship and \'Ecole Normale Superieure (Paris).
The work of VD was supported by Extra-Erasmus Scholarship (Department of Physics, University of Trieste) and Collegio Universitario 'Luciano Fonda'. He also warmly thanks Nordita (Stockholm, Sweden) and KTH (Stockholm, Sweden) for hospitality.

SC and RM acknowledge financial support from the Agence Nationale de la Recherche projects RBMPro (ANR-17-CE30-0021) and Decrypted (ANR-19-CE30-0021).
EA acknowledge the Science for Life Labs (Solna, Sweden) ``Viral sequence evolution research program".

\begin{appendices}

\section{Higher order corrections to the Gaussian closure inference formula}
\label{a:Higher-order-GC}
\setcounter{equation}{0}
\renewcommand\theequation{A.\arabic{equation}}
Starting from from eq. \eqref{eq:DSOC} and exploiting the same argument as in the main body of the paper, it is straightforward to compute  higher order terms in the expansion for $\chi_{ij}$. Let us define for simplicity $\epsilon=1/(4\mu+rc_{ij})$. In the limit $\epsilon\rightarrow0^+$ we write
\begin{equation}\label{eEC}
    \chi_{ij}=\epsilon\chi_{ij}^{(1)}+\epsilon^2\chi_{ij}^{(2)}+\epsilon^3\chi_{ij}^{(3)}+\mathcal{O}(\epsilon^4)\ ,
\end{equation}
and in the case where $f_i = 0$ for all $i$ we find
\begin{align}
    \chi_{ij}^{(1)} &= f_{ij}\\
    \chi_{ij}^{(2)} &= 2\sum_kf_{ik}f_{jk} \\
    \chi_{ij}^{(3)} &= \sum_{k<l} f_{kl}(f_{ik}f_{jl}+f_{jk}f_{il}) -f_{ij}^3 + \sum_{k}\Big[f_{ik}\big(2\sum_{l} f_{kl}f_{jl} - f_{ij} f_{ik}\big)+ f_{jk}\big(2\sum_{l} f_{kl}f_{il} - f_{ij} f_{jk}\big)\Big]
\end{align}
We observe that each correction is of order $L\times \sigma (\{f_{ij}\})$ with respect to the lower one, therefore we expect the expansion to be accurate only if $L\times \sigma (\{f_{ij}\}) \ll 1$.\\
In the more general case where $f_i\ne 0$ for all $i$, we find the first two orders in eq. (\ref{eEC}) to be:
\begin{align}
    \chi_{ij}^{(1)} &= f_{ij}(1-\chi_i^2)(1-\chi_j^2)\label{eFO}\\
    \chi_{ij}^{(2)} &= \sum_k f_{ik}\big(\chi_{jk}^{(1)}+\chi_{ik}^{(1)}\chi_i\chi_j-\chi_i\chi_k\chi_{ij}^{(1)}\big)\ - \sum_{k,l}f_{kl}\chi_i\chi_l\chi_{jk}^{(1)}-\sum_lf_{il}\chi_l\chi_i\chi_{ij}^{(1)}\ + \notag\\
    & + \sum_{k<l} f_{kl} \Big(\chi_{ik}^{(1)}\chi_j\chi_l + \chi_{il}^{(1)}\chi_j\chi_k-2f_i\chi_i\chi_{ij}^{(1)} +  f_{ij}\chi_i\chi_j\chi_{ij}^{(1)}\Big) + \{i\longleftrightarrow j\}
\end{align}
where, for the sake of clarity, in the last equation we have left implicit the terms like $\chi_{ij}^{(1)}$ as specified in eq. (\ref{eFO}).

\section{FFPopSim settings}
\label{a:FFPopSim}
The FFPopSim package, written by
Fabio Zanini and Richard Neher
simulates a population evolving
due to mutation, selection and recombination~\cite{FFPopSim}.

We use the class \texttt{haploid\_highd} i.e.  \emph{individual-based} simulations that handle the population as a set of \emph{clones} $(g_i,n_i(t))$ where $g_i$ is a genotype and $n_i(t)$ is the number of individuals with genotype $g_i$ at time $t$ (only existing clones are tracked). At each generation, the size of each clone is first updated $n_i(t) \rightarrow n_i(t+1) \sim \mathscr{P}_{\lambda}$ where $\mathscr{P}$ is the Poisson distribution with parameter $\lambda = \frac{1}{\langle{e^F}\rangle}e^{F(g_i)+1-\frac{1}{N}\sum_jn_j(t)}$, $N$ is the carrying capacity and $F(g)$ is the fitness function. A fraction $r^*$ (outcrossing rate) of the resulting offspring is destined to the recombination step, paired and reshuffled. Finally, each individual is allowed to mutate with probability $1-e^{-L\mu}$, where $\mu$ is the recombination rate, the exact number of mutations being Poisson distributed $\mathscr{P}_{L\mu}$. \\
We have used FFPoSim in a similar manner as in~\cite{Zeng2020} and we will here only list the settings.
Parameters which are the same
in all simulations reported in this
paper are listed in Tab.~\ref{tab:1}. Parameters that have been varied
(not all variations reported in the paper)
are listed in Tab.~\ref{tab:2}.

It is important to notice that the out-crossing rate $r^*$ in FFPopSim  \textit{a priori} differs from our recombination rate, $r$, appearing in eq.~(\ref{eq:fitness-inference-formula}). In the simulation package, dynamics is discrete in time (with time step of one generation) and $r^*$ is a probability taking value between 0 and 1. In our theory, $r$ is a rate, which can take any positive value. In the examples given in \cite{FFPopSim}, e.g. Fig~2 in the main text and Fig~2 in Supplementary Information the out-crossing probability does not exceed $10^{-2}$. For such low values $r^*$ coincides with a rate (since the time step is equal to unity), which justifies its denomination. We use this correspondence $r^* = 1- e^{-r}\sim r$ between the out-crossing rate $r^*$ in FFPopSim and our recombination rate $r$, valid for small values, to produce the scatter plots in Figs.~\ref{fig:scatter_diff_mu} and \ref{fig:Scatter-plots-diff-sigma}.

Notice that this correspondence breaks down for large recombination rates. Indeed, even for out-crossing rate $r^*=1$ in the simulation package, mutations and fitness effects can still be quite large, depending on the values of the $f_{ij}$'s and of $\mu$, and QLE is not recovered. In the theory, however, all fitness and mutation effects become relatively weak, of the order of $1/r$.

\begin{table}[!ht]
\centering
\caption{\textbf{Main default parameters of FFPopSim used in the simulations.}}
~~\\
\begin{tabular}{ll}
        number of loci (L)              &  25              \\
        number of traits                &   1               \\
        circular                        &  False           \\
        carrying capacity (N)           &   200             \\
        generation                      &   $10,000$               \\
        recombination model             &   CROSSOVERS      \\
        crossover rate   ($\rho$)       &    $0.5$           \\
        fitness additive(coefficients)  &  Gaussian random number with $\sigma(\{f_{i}\})=0.05$  \\
\end{tabular}
\label{tab:1}
\end{table}

\begin{table}[!ht]
\centering
\caption{\textbf{Variable parameters of FFPopSim used in the simulation.}}
~~\\
\begin{tabular}{ll}
    initial genotypes  & binary random numbers\\
    out-crossing rate ($r$) &   $[0. , ~1.0]$  \\
    mutation rate  ($\mu$)  &   $[0.05, ~0.5]$          \\
    epistatic fitness  & Gaussian random number with $\sigma(\{f_{ij}\}) \in$ $[0. 004, ~0.04]$  \\
\end{tabular}
\label{tab:2}
\end{table}

In addition to forward simulations, a subsequent release of the original FFPopSim package allows for the possibility of tracking the genealogy of loci e.g. that of the central locus. Such information can be used in the first place to draw a coalescent tree \cite{Neher2013}: technically, this is done by converting the genealogy in a \texttt{BioPython} tree and using the module \texttt{Bio.Phylo} for plotting purposes.

\begin{figure}[!ht]
\centering
\includegraphics[width=0.8\textwidth]{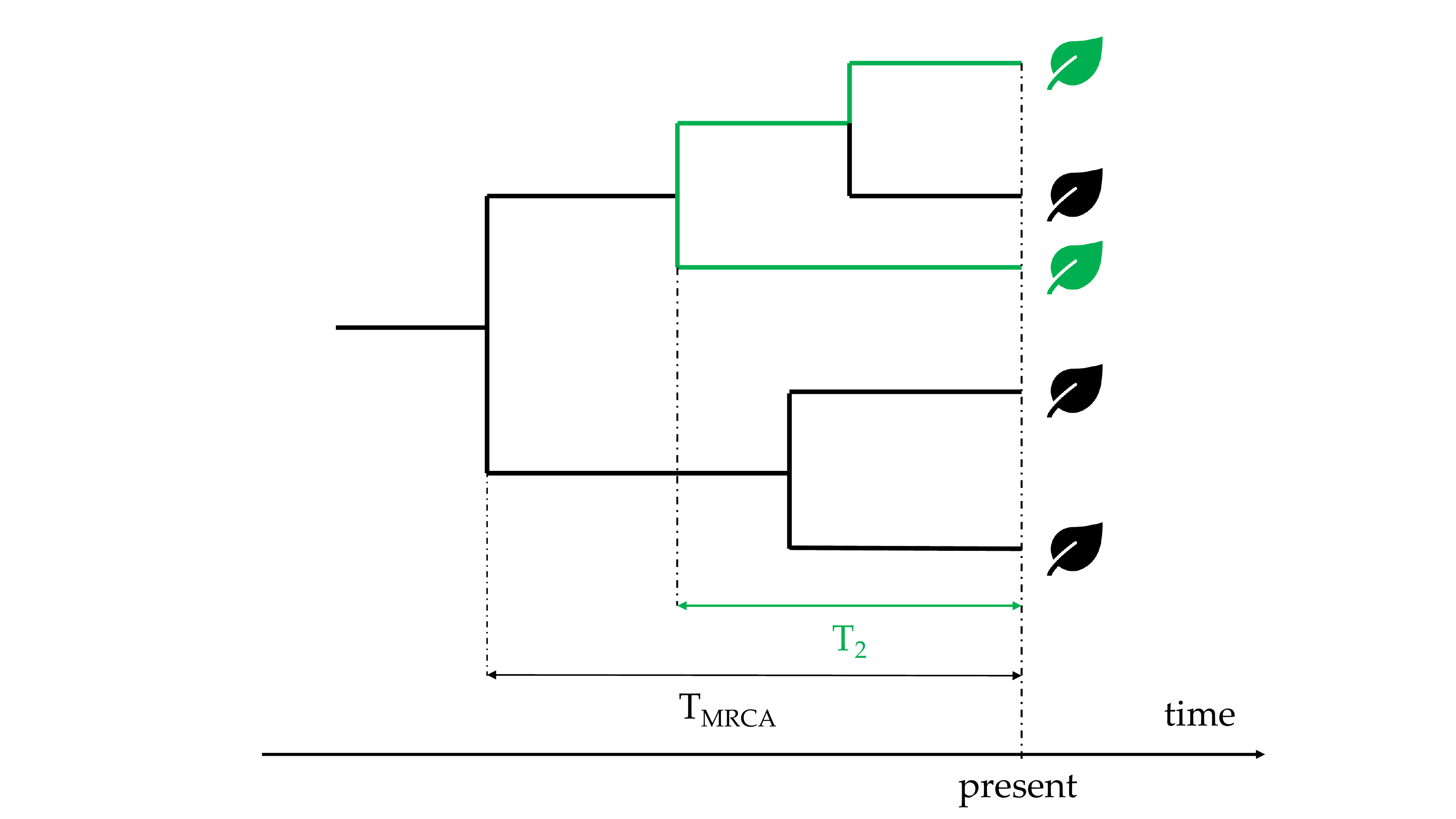}
\caption{Illustrative coalescent tree. The Time to the Most Recent Common Ancestor $T_{MRCA}$ of a tree is the difference between the current time and the time point where all the branches merge. The pair coalescent time $T_2$ for two chosen leaves (individuals) is the $T_{MRCA}$ with respect to their subtree (highlighted in green).}
\label{fig:T2_illustration}
\end{figure}

A quantitative analysis of such trees can be carried out, for instance the Time to the Most Recent Common Ancestor $T_{MRCA}$ of a group of individuals at time $t$ is nothing but the temporal distance of the leaves (individuals) from the root (common ancestor) in the corresponding coalescent tree $\mathcal{CT}$, as shown in Fig. \ref{fig:T2_illustration}. In the same vein, we are able to evaluate the average pair coalescent time $\langle T_2 \rangle$: we sample $n_{2}=10$ pairs of leaves, per each of them we extract the information about their subtree $\mathcal{CT}_2$ and evaluate the $T_{MRCA}(\mathcal{CT}_2)$, which now corresponds to the difference between the present and the time in the past when the two branches stemming from the chosen leaves merge. Averaging over the sample of size $n_2$ gives an estimate of the desired quantity.

\section{Naive mean-field (nMF)}
\label{a:nMF}
Naive mean-field is based on
mimizing the reverse Kullback-Leibler distance
between an empirical probability distribution
and a trial distribution in the family
of independent (factorized) distributions.
This leads to the \textit{inference formula}
$J^{*,nMF}_{ij} =\left(\chi^{-1}\right)_{ij}$.
If the correlation $\chi_{ij}$
is computed as an average over the population
at a single time we call it \textit{single-time-nMF}.
If on the other hand $\chi_{ij}$
is computed by additionally averaging
over time we call it \textit{all-time-nMF}.

The pseudo-code for nMF inference
taking $\chi_{ij}$ as input
is presented in Algorithm \ref{alg:nMF}.
\begin{algorithm}\label{alg:nMF}
\LinesNumbered
\caption{Epistatic fitness inference by KNS formula \eqref{eq:original_KNS_equation} 
with $J_{ij}^*$ reconstructed by nMF procedure: $f_{ij}^{nMF}$}
 \SetAlgoLined
 \setstretch{1.35}
   \KwIn{mean correlations: $\left<\chi_{ij}\right>$}
   \KwOut{inferred epistatic fitness: $f_{ij}^{nMF}$ }
\begin{algorithmic}[1]
       \STATE import \textbf{scipy}
       \STATE from \textbf{scipy} import \textbf{linalg}
       \STATE $J_{ij}^{nMF}$ = - linalg.inv($\left<\chi_{ij}\right>$)
       \STATE $f_{ij}^{nMF} = J_{ij}^{nMF} \ast  r c_{ij}$
\end{algorithmic}
\end{algorithm}

\section{Numerical comparison between eq. (\ref{eq:low-recomb-_KNS_equation}) and eq. (\ref{eq:fitness-inference-formula})}
\label{a:fitness-dca-chi}
\begin{figure*}[!ht]
\centering
\subfigure{
\begin{minipage}[t]{0.31\linewidth}
\centering
\includegraphics[width=\textwidth]{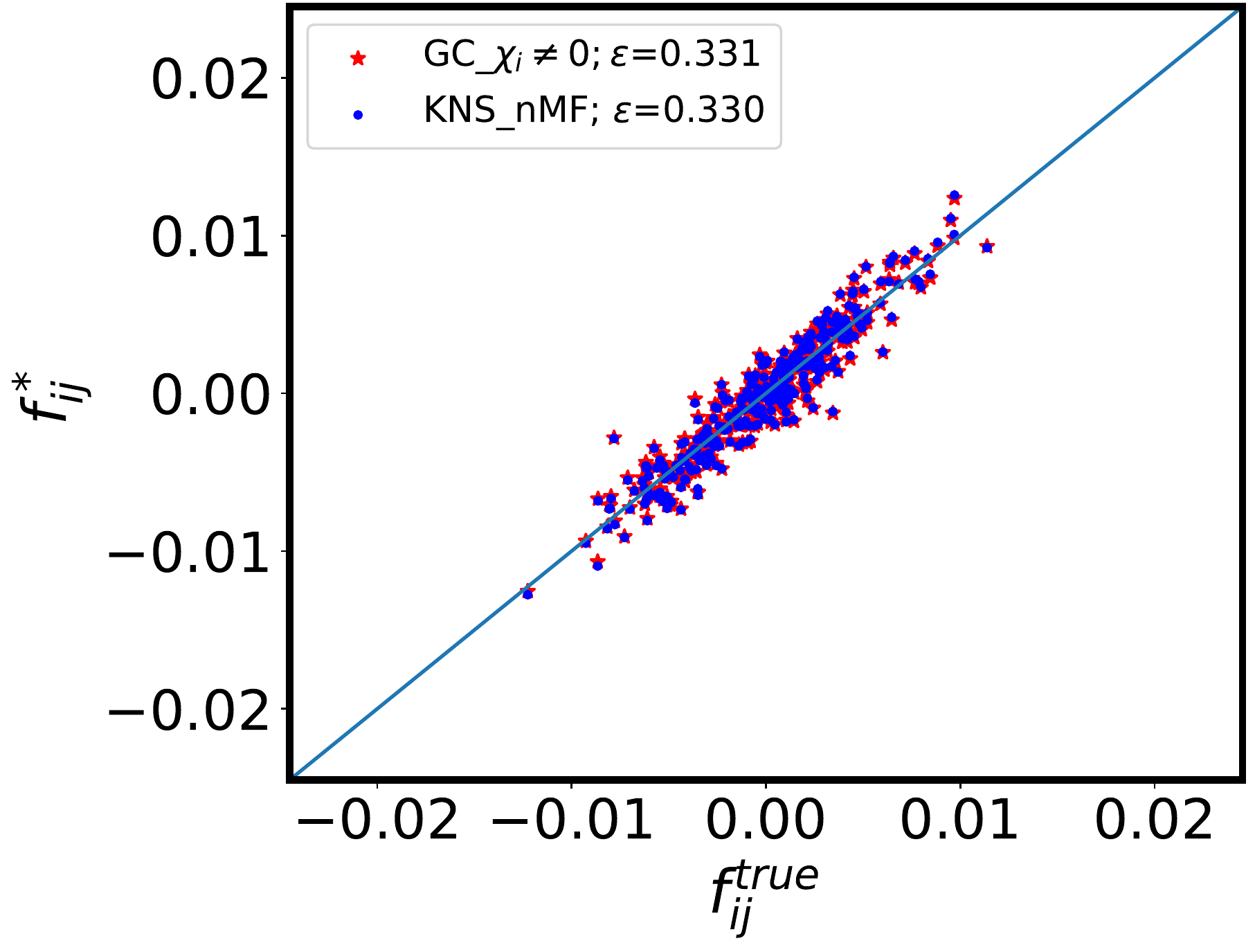}
\put(-101,-11){\small (a)~$\sigma=0.004, r=0.2$}
\end{minipage}%
}
\subfigure{
\begin{minipage}[t]{0.31\linewidth}
\centering
\includegraphics[width=\textwidth]{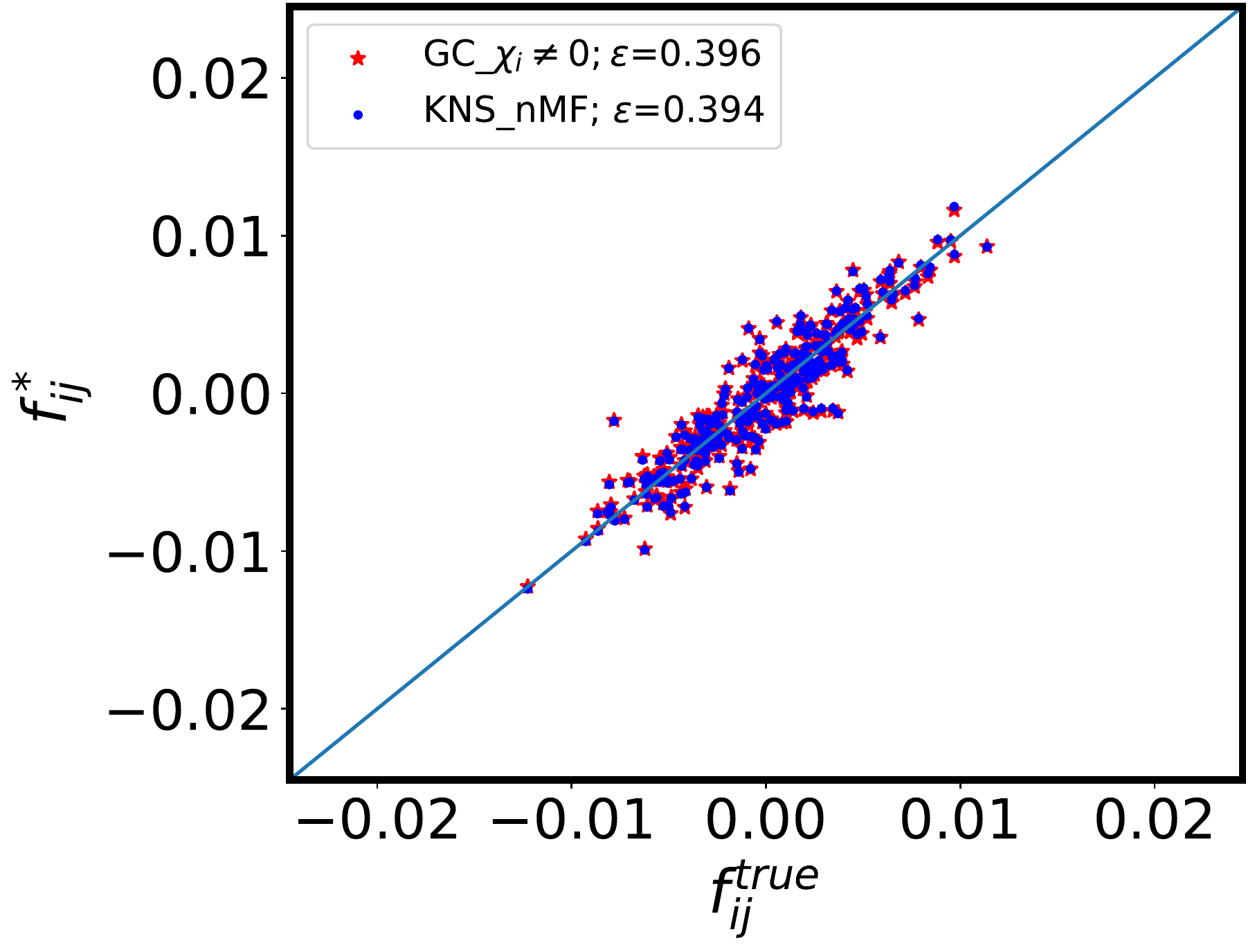}
\put(-101,-11){\small (b)~$\sigma=0.004, r=0.5$}
\end{minipage}%
}
\subfigure{
\begin{minipage}[t]{0.31\linewidth}
\centering
\includegraphics[width=\textwidth]{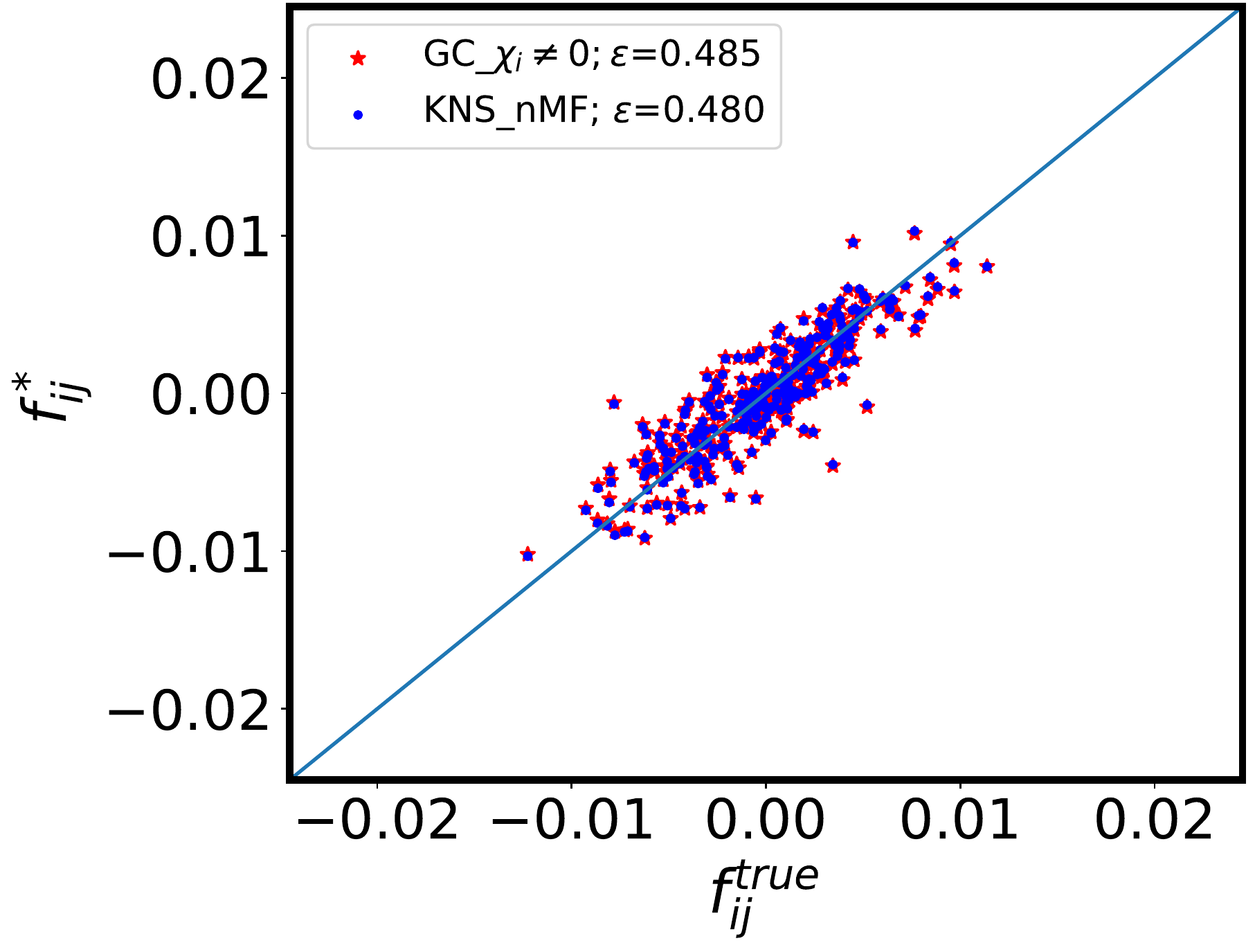}
\put(-101,-11){\small (c)~$\sigma=0.004, r=0.8$}
\end{minipage}%
}\\
\subfigure{
\begin{minipage}[t]{0.31\linewidth}
\centering
\includegraphics[width=\textwidth]{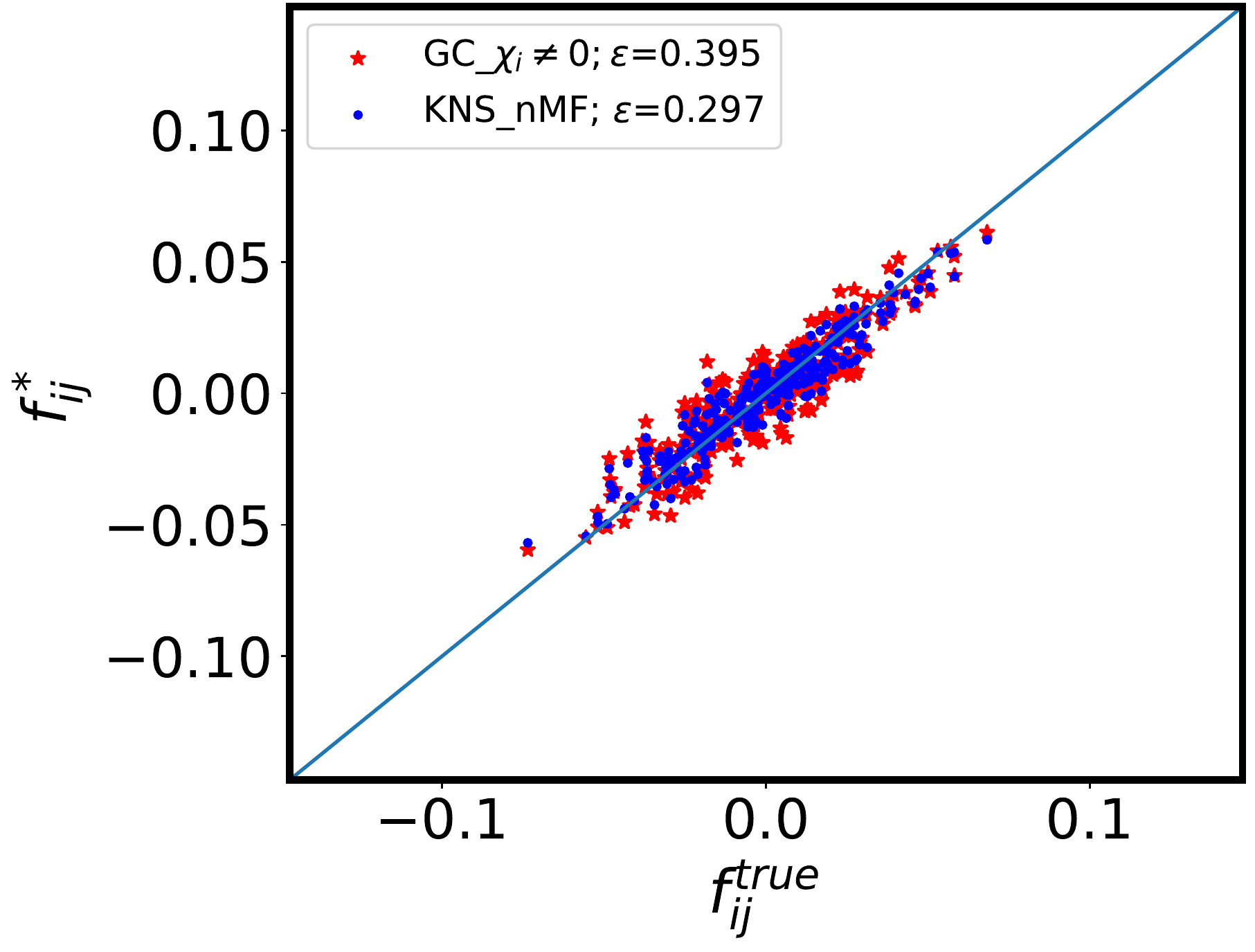}
\put(-101,-11){\small (d)~$\sigma=0.024, r=0.2$}
\end{minipage}%
}
\subfigure{
\begin{minipage}[t]{0.31\linewidth}
\centering
\includegraphics[width=\textwidth]{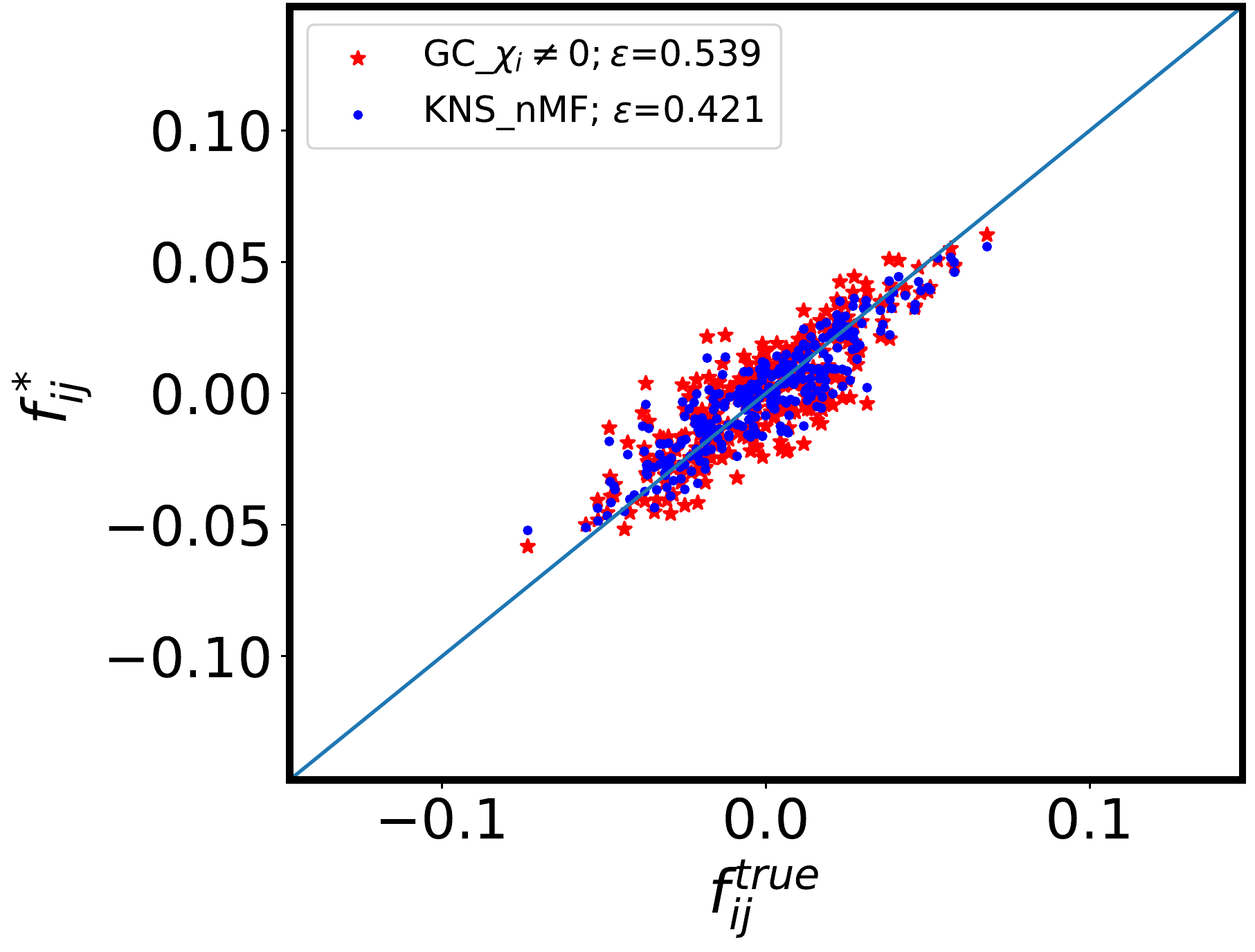}
\put(-101,-11){\small (e)~$\sigma=0.024, r=0.5$}
\end{minipage}%
}
\subfigure{
\begin{minipage}[t]{0.31\linewidth}
\centering
\includegraphics[width=\textwidth]{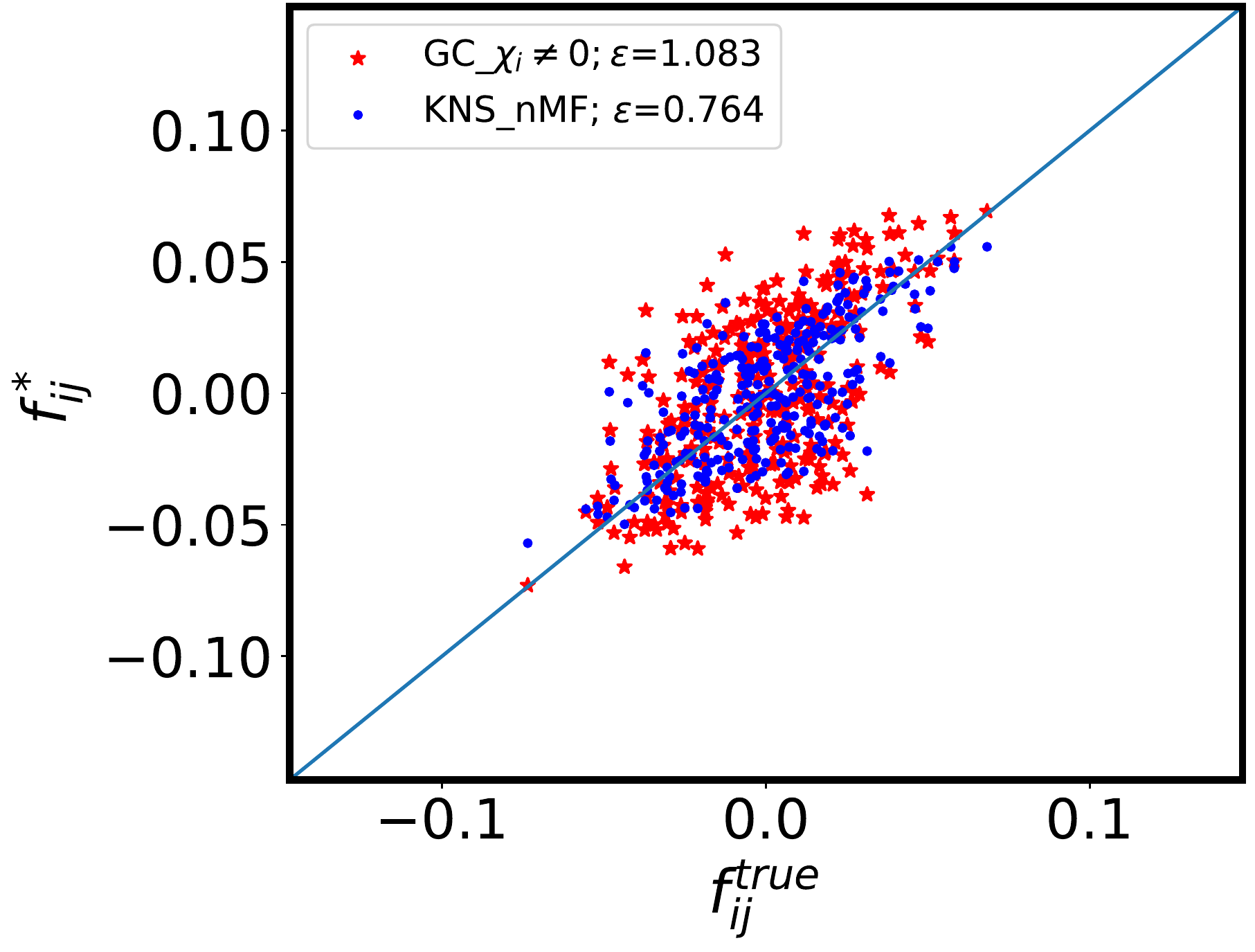}
\put(-101,-11){\small (f)~$\sigma=0.024, r=0.8$}
\end{minipage}%
}\\
\subfigure{
\begin{minipage}[t]{0.31\linewidth}
\centering
\includegraphics[width=\textwidth]{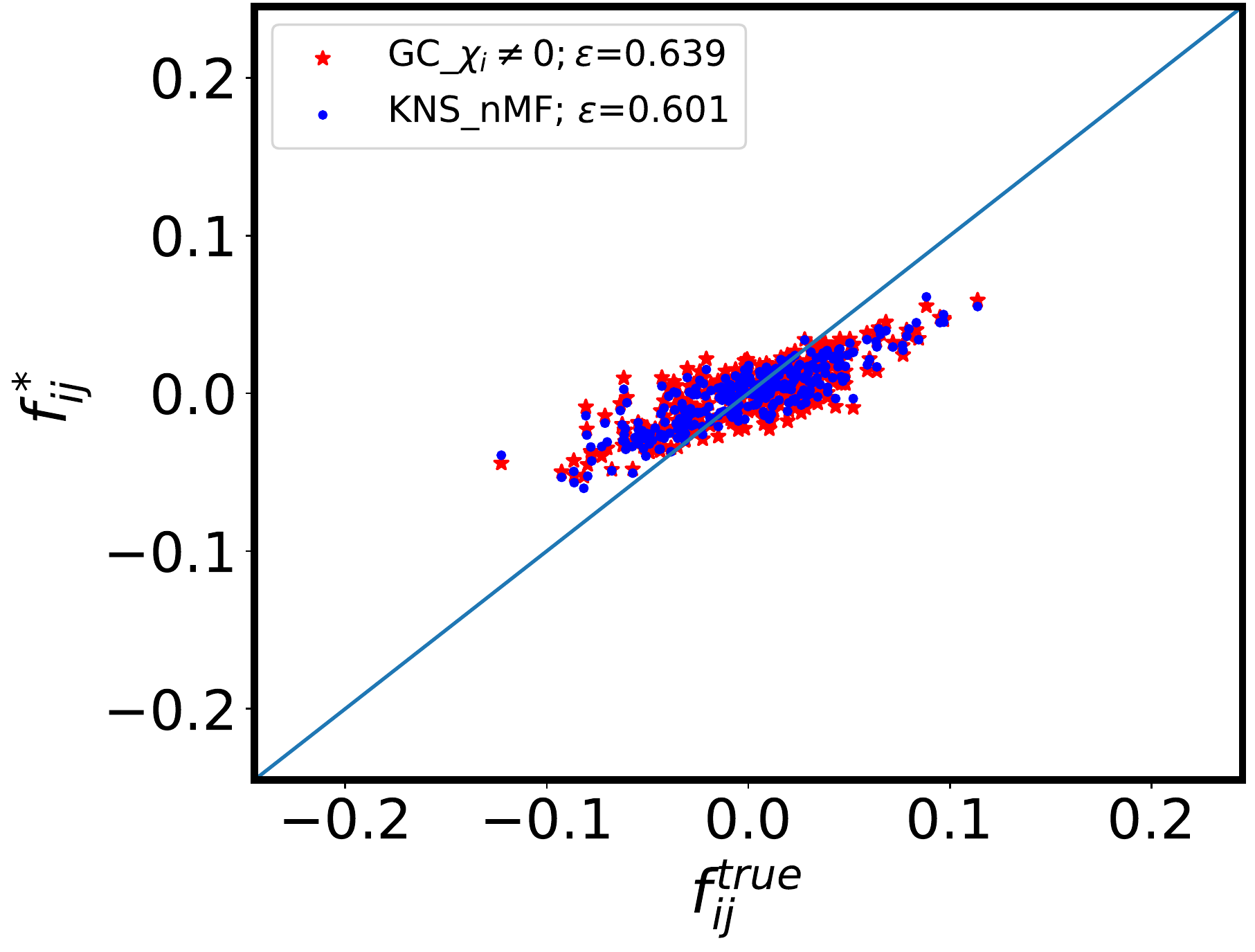}
\put(-101,-11){\small (g)~$\sigma=0.04, r=0.2$}
\end{minipage}%
}
\subfigure{
\begin{minipage}[t]{0.31\linewidth}
\centering
\includegraphics[width=\textwidth]{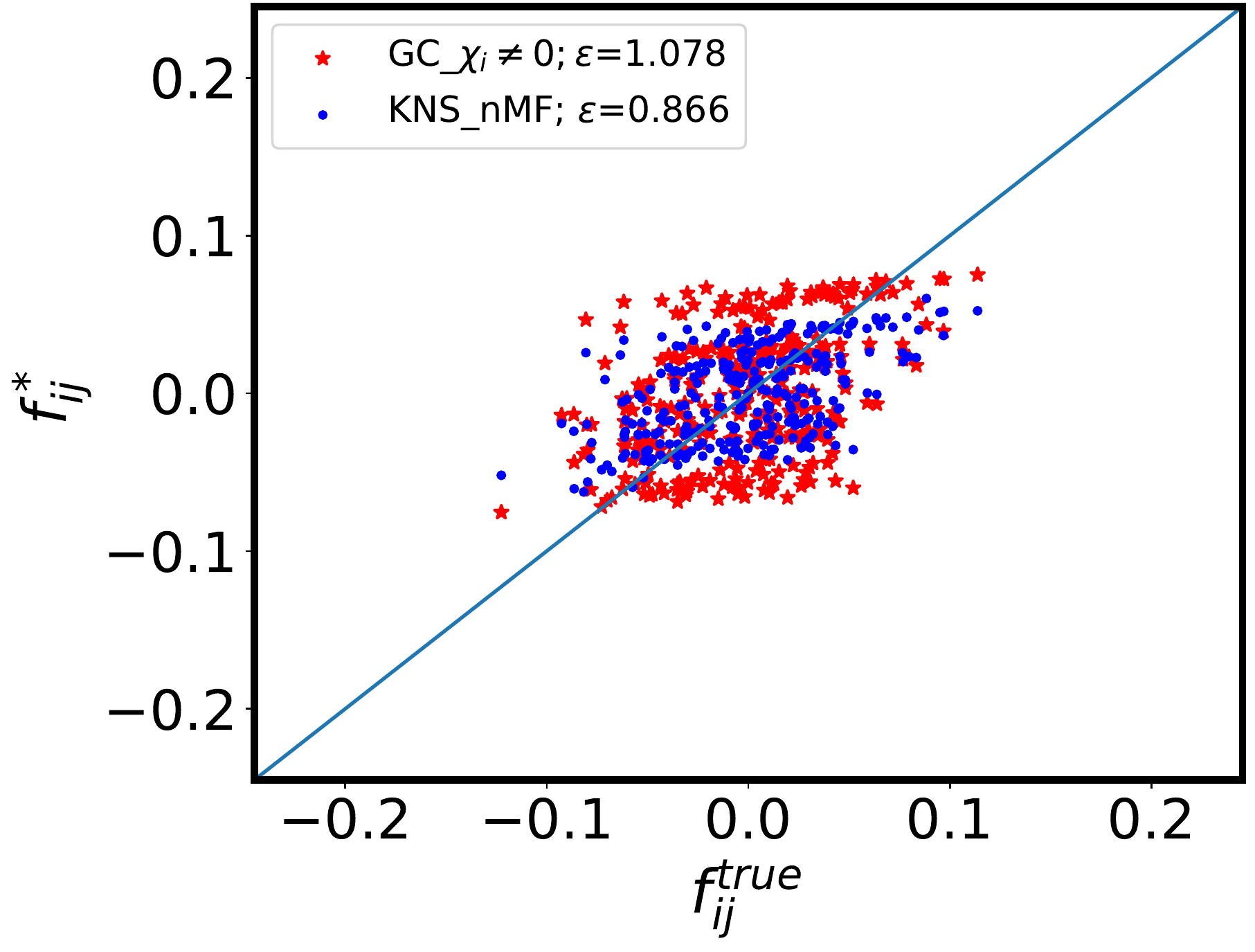}
\put(-101,-11){\small (h)~$\sigma=0.04, r=0.5$}
\end{minipage}%
}
\subfigure{
\begin{minipage}[t]{0.31\linewidth}
\centering
\includegraphics[width=\textwidth]{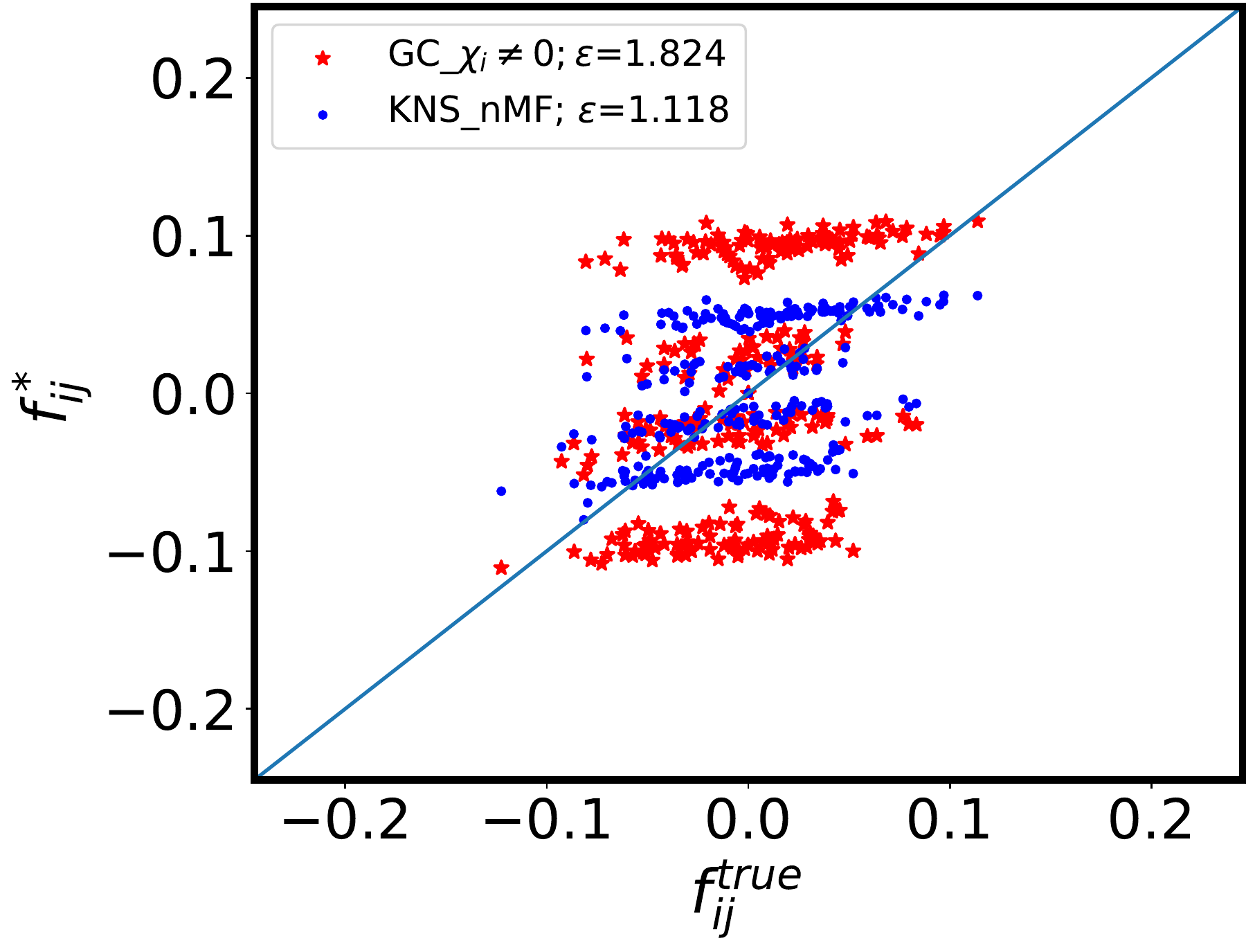}
\put(-101,-11){\small (i)~$\sigma=0.04, r=0.8$}
\end{minipage}%
}
\caption{Scatter plots for testing and reconstructed $f_{ij}$s. The standard deviation $\sigma(\{f_{ij}\}^{true})$ increases from top to bottom rows ($0.004, 0.024$ and $0.04$ respectively) and recombination rate $r$ enlarges in columns from left to right ($0.2, 0.5$ and $0.8$ respectively). Red stars for $f_{ij}^* = \chi_{ij} \cdot (4\mu + rc_{ij})/((1-\chi_i^2)(1-\chi_i^2))$ and blue dots for $f_{ij}^*=(4\mu + rc_{ij}) \cdot J_{ij}^{*,nMF}$.
The other parameters are the same to those in Fig. \ref{fig:Scatter-plots-diff-sigma}.
In the regime of weak $\sigma$ and $r$, the reconstructions are equivalent. Increasing $\sigma$ for sufficiently small $r$ as in (d),(e),(g) the mean field reconstruction outperforms the Gaussian one. However, both reconstructions fail for sufficiently high $\sigma,r$, as in (f),(g),(h),(i), where strong correlations emerge between loci that drive the system out of the QLE phase \cite{Dichio2020,DZA}. One realization of the fitness terms $f_{ij}$ and $f_i$ for each parameter value.}
\label{fig:app-Scatter-plots-diff-sigma}
\end{figure*}

To compare the results of epistasis inference by eq. (\ref{eq:low-recomb-_KNS_equation}) from KNS theory and eq. (\ref{eq:fitness-inference-formula}) through Gaussian closure, we present the numerical simulations in Fig. ~\ref{fig:app-Scatter-plots-diff-sigma} with a fixed mutation rate $\mu=0.2$ while different $\sigma(\{f_{ij}\})$s and recombination rate $r$.
The blue dots are for eq. \eqref{eq:low-recomb-_KNS_equation} while the red stars for eq. \eqref{eq:fitness-inference-formula}.
As shown in the top row of Fig. \ref{fig:app-Scatter-plots-diff-sigma} (a), (b) and (c), two methods perform almost the same for weak epistatic fitness $\sigma(\{f_{ij}\})=0.004$.
When increasing $\sigma(\{f_{ij}\})$ and for sufficiently low recombination rates as in Fig. \ref{fig:app-Scatter-plots-diff-sigma} (d), (e), (g) , we observe that \eqref{eq:low-recomb-_KNS_equation} works considerably better than eq.~\eqref{eq:fitness-inference-formula}, as it is evident from the smaller reconstruction error of the former with respect to the latter.
Finally, none of them works for large $\sigma(\{f_{ij}\})$ and high $r$ as shown in Fig.~\ref{fig:app-Scatter-plots-diff-sigma}~(f), (h) and (i). The parameters for these cases are located in the white area of Fig.~\ref{fig:figs-sigma-r} where the system may not be in the QLE state and both the reconstructions (Neher-Shraiman and Gaussian closure) fail. This part with strong correlations has been studied extensively in VD's Master's thesis ~\cite{Dichio2020,DZA}.

\section{Effects of Genetic Drift}
\label{a:Genetic Drift}
\begin{figure}[!ht]
        \centering
        \includegraphics[width=0.5\textwidth]{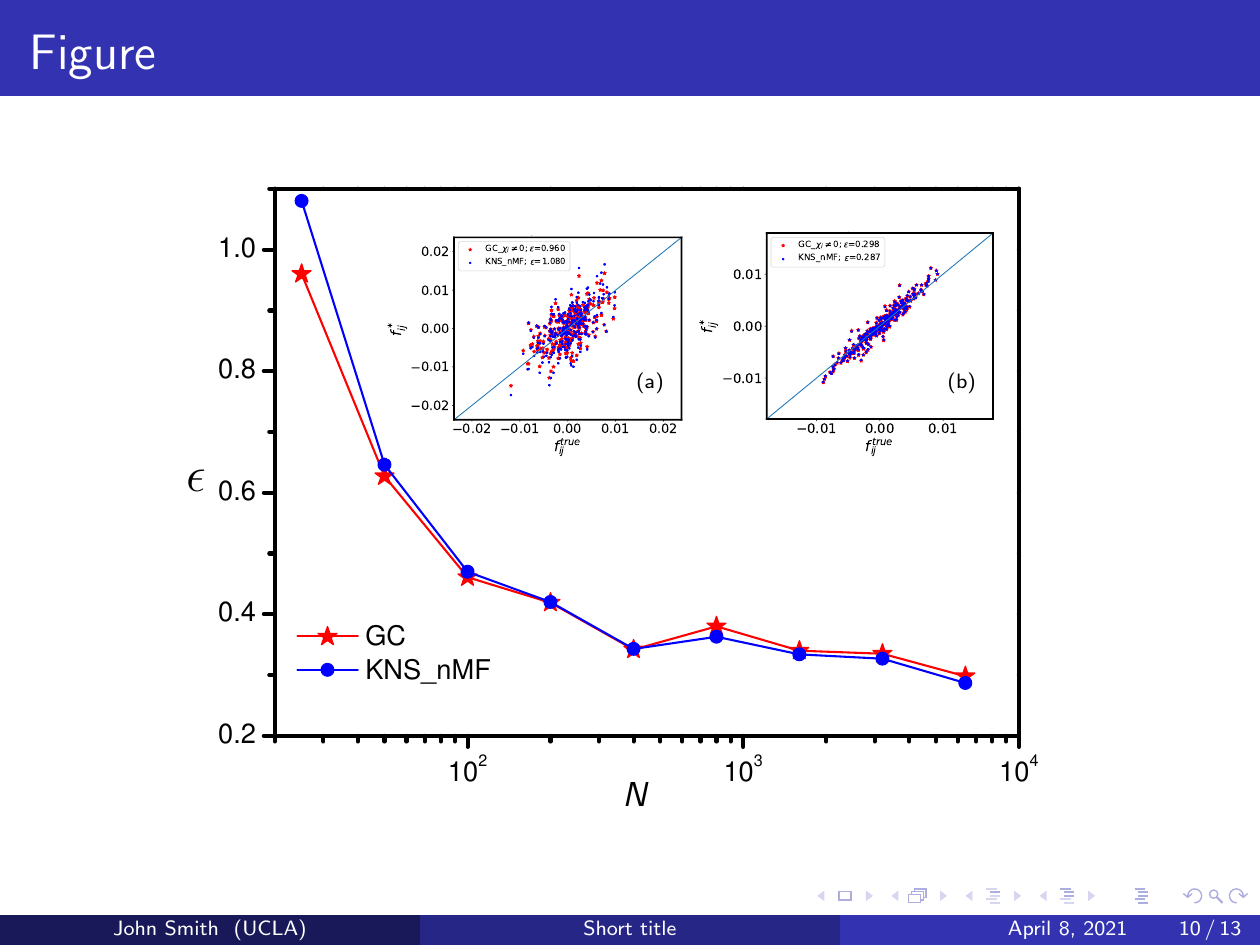}
        \caption{Semi-log plot for epistasis reconstruction error $\epsilon$ versus the average size of population $N$. Insert (a): scatter plot for testing and reconstructed $f_{ij}$s with $N=L=25$; Insert (b): scatter plot with $N=6400$.  Red stars for $f_{ij}^*=\chi_{ij}\cdot(4\mu+rc_{ij})/((1-\chi_i^2)(1-\chi_j^2))$ and blue dots for  $f_{ij}^*=J_{ij}^{*,nMF}\cdot(4\mu+rc_{ij})$. Epistasis $f_{ij}$ are recovered roughly better with increasing $N$.
        The other parameter values:  $\sigma(f_i) = 0.05$, $\sigma(f_{ij}) = 0.004$, mutation rate $\mu = 0.25$, out-crossing rate $r=0.5$, cross-over rate $\rho= 0.5$, number of loci $L = 25$, carrying capacity $N = 200$, generations $T = 10,000$. One realization of the fitness terms $f_{ij}$ and $f_i$ for each parameter value. }
        \label{fig:epsilon_N}
    \end{figure}

The effects of genetic drift on epistasis effects are studied through the inference error $\epsilon$ with different population sizes $N$. It is presented in a semi-log plot as shown in the main panel of Fig.~\ref{fig:epsilon_N}. The red stars are for the epistasis inference error given by eq.~(\ref{eq:fitness-inference-formula})~ $f_{ij}^*=\chi_{ij}\cdot(4\mu+rc_{ij})/((1-\chi_i^2)(1-\chi_j^2))$ while blue dots for  eq.~(\ref{eq:low-recomb-_KNS_equation}) $f_{ij}^*=J_{ij}^{*,nMF}\cdot(4\mu+rc_{ij})$. 
There is a clear trend that both methods work better with increasing population sizes. However, eq.~(\ref{eq:fitness-inference-formula}) works slightly better when the population size is less than 400 while eq.~(\ref{eq:low-recomb-_KNS_equation}) recovers the epistasis better when $N>400$.  The inserts (a) and (b) of Fig.~\ref{fig:epsilon_N} show the scatter plots for the recovered and testing epistasis $f_{ij}$s with $N=25$ (equal number with that of locus in an individual sequence) and $N=6400$ respectively. Clearly both eqs. recover the epistasis better with large population size compared with that with small ones
. 

\section{Epistasis inference with directional selections}
\label{a:Gaussian_additive_effects}
\begin{figure}[!ht]
    \centering
    \includegraphics[width=0.5\textwidth]{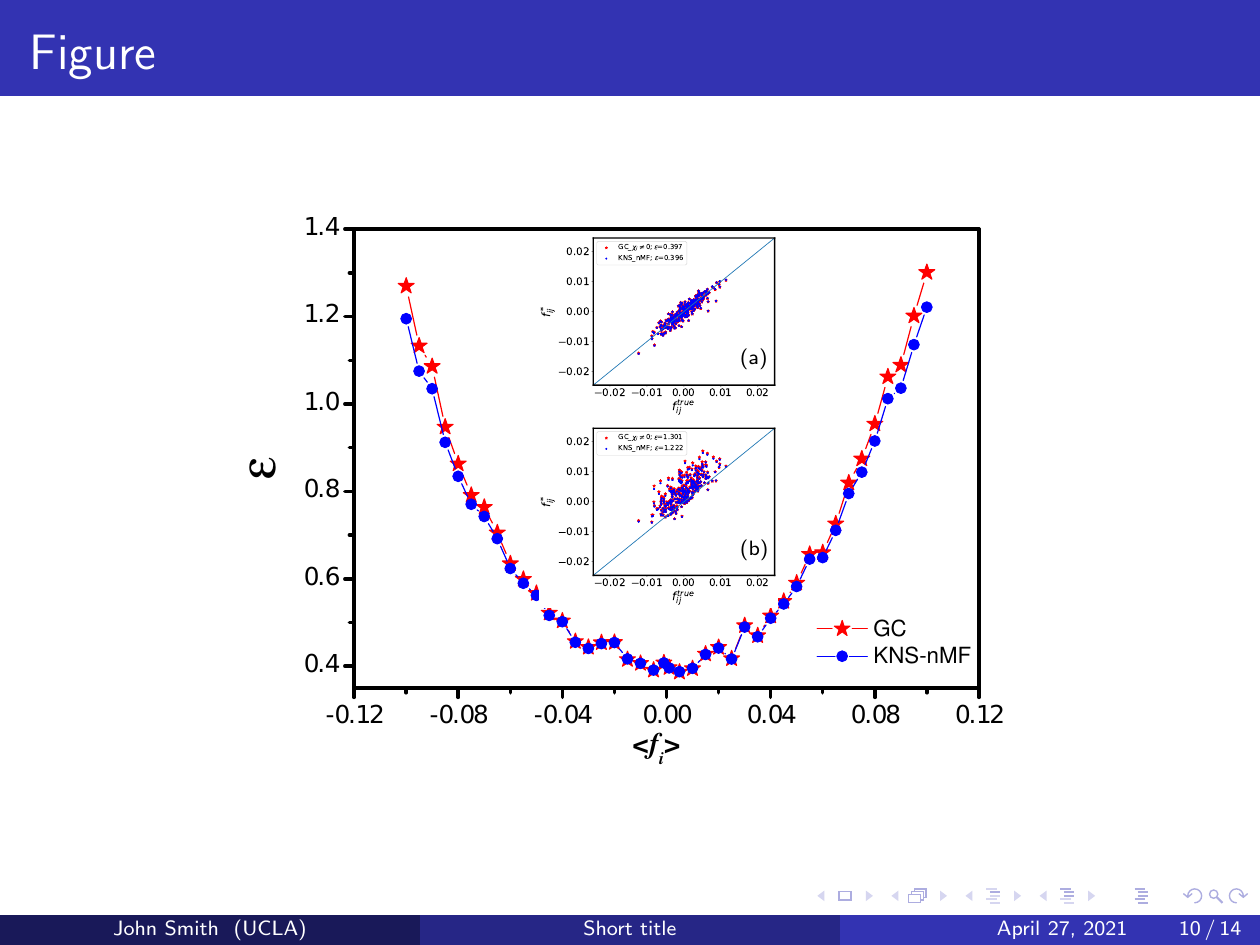}
    \caption{Epistasis reconstruction error $\epsilon$ versus the means of Gaussian distributed additive fitness $\langle f_i\rangle$. Insert (a): scatter plot for testing and reconstructed $f_{ij}$s with $\langle f_i\rangle=0.001$; Insert (b): scatter plot with $\langle f_i\rangle=0.01$.  Red stars for $f_{ij}^*=\chi_{ij}\cdot(4\mu+rc_{ij})/((1-\chi_i^2)(1-\chi_j^2))$ and blue dots for  $f_{ij}^*=J_{ij}^{*,nMF}\cdot(4\mu+rc_{ij})$. The epistasis reconstructions are getting worse with stronger directional fields.      The other parameter values: standard deviation $\sigma(\{f_{ij}\}) = 0.004$, mutation rate $\mu = 0.25$, out-crossing rate $r=0.5$, cross-over rate $\rho= 0.5$, number of loci $L = 25$, carrying capacity $N = 200$, number of generations $T = 10,000$. One realization of the fitness terms $f_{ij}$ and $f_i$ for each parameter value.  }
    \label{fig:epsilon_ave_fi}
\end{figure}

This appendix summarizes the effects of non-zero additive fitness on epistasis inference through numerical simulations. Here the additive effects $f_i$s are Gaussian distributed with non-zero means and the standard deviations are fixed as $\sigma(\{f_i\})=0.05$. The red stars for the epistasis inference with Gaussian closure eq.~(\ref{eq:fitness-inference-formula})  while the blue dots for the revised KNS method by eq.~(\ref{eq:low-recomb-_KNS_equation}). The inserts of Fig.~\ref{fig:epsilon_ave_fi} show the scatter plots for the recovered and testing epistasis effects with (a): $\langle f_i\rangle=0.001$ and (b): $\langle f_i\rangle=0.01$ respectively. The other parameters for each points in the main panel are as follows: standard deviation of the pairwise epistasis fitness $\sigma(\{f_{ij}\}) = 0.004$ and that of the single-locus additive fitness $\sigma(\{f_i\}) = 0.05$, mutation rate $\mu = 0.25$, out-crossing rate $r=0.5$, cross-over rate $\rho= 0.5$, number of loci $L = 25$, carrying capacity $N = 200$, generations $T = 10,000$.


Both methods recover the tested epistasis better with weaker means of additive fitness compared with that following stronger directional selections. It is notable that the reconstructed epistasis have a roughly corrected trends with large additive fitness, as shown in the inner panel (b) of Fig.~\ref{fig:epsilon_ave_fi} for $\langle f_i \rangle=0.01$. This may indicates the revision of the epistasis inference formulae in our work for stronger directional selections.

\end{appendices}

\newpage
\bibliographystyle{iopart-num}

\bibliography{ref,ref2,ref3}

\end{document}